\documentclass[%
 aip,
 amsmath,amssymb,
 reprint,%
]{revtex4-1}

\usepackage{graphicx}
\usepackage{dcolumn}
\usepackage{bm}
\usepackage[usenames,dvipsnames]{color}
\usepackage{color,linegoal}
\usepackage[dvipsnames]{xcolor}
\definecolor{myc}{rgb}{0,0,0}

\usepackage[utf8]{inputenc}
\usepackage[T1]{fontenc}
\usepackage{mathptmx}
\usepackage{etoolbox}
\usepackage{float}
\usepackage{hhline}

\makeatletter
\def\@email#1#2{%
 \endgroup
 \patchcmd{\titleblock@produce}
  {\frontmatter@RRAPformat}
  {\frontmatter@RRAPformat{\produce@RRAP{*#1\href{mailto:#2}{#2}}}\frontmatter@RRAPformat}
  {}{}
}%
\makeatother
\begin{document}

\newcommand{\aAA}{\ensuremath{\alpha_{\scriptscriptstyle AA}}}
\newcommand{\aBB}{\ensuremath{\alpha_{\scriptscriptstyle BB}}}
\newcommand{\aAB}{\ensuremath{\alpha_{\scriptscriptstyle AB}}}
\newcommand{\aAP}{\ensuremath{\alpha_{\scriptscriptstyle AP}}}
\newcommand{\aBP}{\ensuremath{\alpha_{\scriptscriptstyle BP}}}
\newcommand{\sAA}{\ensuremath{\sigma_{\scriptscriptstyle AA}}}
\newcommand{\sBB}{\ensuremath{\sigma_{\scriptscriptstyle BB}}}
\newcommand{\sAB}{\ensuremath{\sigma_{\scriptscriptstyle AB}}}
\newcommand{\sPP}{\ensuremath{\sigma_{\scriptscriptstyle PP}}}
\newcommand{\epsAB}{\ensuremath{\varepsilon_{\scriptscriptstyle AB}}}
\newcommand{\epsAA}{\ensuremath{\varepsilon_{\scriptscriptstyle AA}}}
\newcommand{\epsBB}{\ensuremath{\varepsilon_{\scriptscriptstyle BB}}}

\preprint{AIP/123-QED}

\title[]{Mixing-demixing transition and void formation in quasi-2D binary mixtures on a sphere}

\author{D. Truzzolillo}
\email{domenico.truzzolillo@umontpellier.fr}
\affiliation{Laboratoire Charles Coulomb (L2C), UMR 5221 CNRS-Universitè de Montpellier, F-34095 Montpellier, France}

\date{\today}

\begin{abstract}
Motivated by observations of heterogeneous domain structure on the surface of cells and vesicles and by domain formation due to the adsorption of complex molecules onto composite membranes, we consider a minimal quasi 2D-model to describe the structure of binary mixtures on the surface of a spherical particle. 
We study the effect of miscibility and adsorbing particle (AP) addition on the mixture structure. We define a new scalar quantity, the geodesic mixing parameter $\Xi$, through which we detail the effect of miscibility and the role of preferential affinity of APs with one of the two components of the mixture, distinguishing unambiguously between mixing and demixing solely induced by APs.  
Finally, by inspecting the distributions of void sizes, we show how void formation is ruled by miscibility and AP-mixture interactions, which control the transition from exponentially-tailed to fat-tailed distributions.\footnote{The following article has been accepted by \emph{The Journal of Chemical Physics}. After it is published, it will be found at https://aip.scitation.org/journal/jcp}
\end{abstract}

\maketitle

\section{\label{Intro}Introduction}
The investigation of the equilibrium properties
of particles whose motion is confined to curved surfaces is surely challenging, but it is essential to achieve a satisfactory understanding of processes such as the lateral diffusion of proteins in membranes \cite{malchusAnomalousDiffusionReports2010,javanainenAnomalousNormalDiffusion2013,metzlerNonBrownianDiffusionLipid2016}, the controlled synthesis of colloidal aggregates\cite{savaralaStabilizationSoftLipid2011}, and demixing in lipid bilayers\cite{andersonRoleLipidShells2002,binderDomainsRaftsLipid2003,veatchCriticalFluctuationsPlasma2008}. 
Confining particles to the surface of a sphere introduces
several constraints to their placement that are absent in Euclidean
(flat) space. Both the curvature and the topology of
the sphere play a role in the way particles can be arranged
on it—most prominently, no regular lattice can be fit onto
its surface, as the topology of the sphere requires \textcolor{myc}{the presence of 12 pentagonal disclinations \cite{bowickTwodimensionalMatterOrder2009}, and introduces structural defects characterizing glassy fluids at large area fractions \cite{singhCooperativelyRearrangingRegions2020}}. The compactness of the sphere also requires a careful treatment: instead of the straightforward thermodynamic limit, configurations of finite
(small) numbers of particles become more relevant, and both
the number of particles and the size of the sphere need to be
considered as two independent parameters \cite{postStatisticalThermodynamicsParticles1986}.
Among all possible spherical systems, multicomponent mixtures are of particular importance since they represent the class of systems including biological membranes in which the self-organisation of lipids into different domains is critical for many cell properties such as recognition, signalling, or endocytosis and exocytosis \cite{salaunLipidRaftsRegulation2004}. 
Demixing and segregation in bulk systems can proceed via a number
of distinct physical mechanisms (such as spinodal decomposition
and nucleation), and it is generally well understood. However, the impact of spatial confinement and the interaction with external agents on the structure of multicomponent mixtures are less clear. Confinement can arise from the presence of external fields, produced, for example,
by substrates or random obstacles, but they can also be imposed
by the geometry of the embedding space. The latter type of
confinement is particularly relevant to biological cells, for
which the mobile fluid particles constituting the cell membrane
are constrained to lie on the surface of a quasi-spherical
body. \textcolor{myc}{This has boosted the interest for a deep understanding of multiphase fluids on smooth manifolds, especially on spheres and spheroids, and so in the past 2 decades many numerical studies have been focused on the particular problem of the phase separation of binary systems on surfaces and its dependence on the surface curvature. \cite{tangPhaseSeparationPatterns2005,greerFourthOrderPartial2006,marenduzzoPhaseSeparationDynamics2013,barreiraSurfaceFiniteElement2011}
Biological membranes} are indeed described as a quasi two-
dimensional fluids \cite{saffmanBrownianMotionBiological1975, camleyDynamicScalingPhase2011} in which the different constituents flow within the membrane manifold, being still partially free to move orthogonally to the membrane tangent plane and coupled to a less viscous three-dimensional surrounding medium.
They exhibit very often stable domains, whose spatial
distribution is important for many cell properties\cite{lennePhysicsPuzzlesMembrane2009,lingwoodLipidRaftsMembraneOrganizing2010}. 
One common view is that the domains on the surface of
a cell are a consequence of an arrested or incomplete phase
separation, however it remains to be established whether the
observed states are permanent or metastable in character.
If the heterogeneous domain structure on the cell surface
is an equilibrium state, then some stabilizing mechanism
is required; the line tension present at interfaces between
domains would make inhomogeneous phases energetically
unfavorable when compared to a fully phase-separated system.


Many works have suggested that biological membranes are close to a demixing line in the temperature-composition plane. 
When the lipid mixture is close to the miscibility critical point\cite{onukiPhaseTransitionDynamics2002,honerkamp-smithIntroductionCriticalPoints2009a} (the so-called weak-segregation limit) the temperature $T$ is close to a critical temperature $T_c$. In this limit, density fluctuations give rise to wide interfaces between fluctuating lipid domains, which do not have a precise shape\cite{destainvilleRationaleMesoscopicDomain2018}.

As a matter of fact, if a membrane is composed of two or more lipids (as most
biomembranes are), the segregation of one of the
lipid components in a set of domains is often encountered, as for example rafts rich in cholesterol and sphingolipids that have been implicated recently in signal transduction and membrane trafficking pathways\cite{andersonRoleLipidShells2002,binderDomainsRaftsLipid2003}.
Veatch and coworkers\cite{veatchCriticalFluctuationsPlasma2008} suggested
that lateral heterogeneities in living cells at physiological conditions correspond to critical fluctuations and that perturbations that alter the phase boundary could have a large effect on the size, composition, and
lifetime of fluctuations at physiological
temperatures. Also, lipid demixing due to the existence of a miscibility critical point in red blood cell (RBC) membranes has been highlighted via Langmuir trough and epifluorescence experiments by Keller and coworkers \cite{kellerRedBloodCell1998}, who showed that monolayers of lipids extracted from RBC membranes at room temperature segregate and form domains above a threshold surface pressure at the air-water interface. 

In addition to single membrane properties, the interaction potential between
a pair of cells is also strongly influenced by the distribution and
size of the domains covering its surface. In this sense, cells
may be regarded as a naturally occurring type of “patchy
particles,” which are particles with distinct
surface sites generating anisotropic interparticle interactions.
While synthetically fabricated patchy particles have attractive
patches strategically arranged on their surface
\cite{bianchiPatchyColloidsState2011}, the domains covering a cell emerge as a result of self-organization. 
Recent developments in the controlled fabrication of patchy particles have raised hopes that materials with desired properties may be tailored by prescribing the number and geometrical arrangement of the patches \cite{bianchiPatchyColloidsState2011}.
In this framework, confined binary mixtures of active and passive particles on a sphere have also been studied\cite{aiBinaryMixturesActive2020} very recently, since particle activity coupled to the presence of intrinsic surface curvature frustrates local order, giving rise to novel phenomena \cite{shankarTopologicalSoundFlocking2017,keberTopologyDynamicsActive2014,sanchezSpontaneousMotionHierarchically2012}, such as curvature-induced defects and unforced flow.    

On top of the spontaneous formation of domains and defects on spheres, both loss of miscibility and domain destabilization in spherically confined fluids can be also caused by the addition of adsorbing agents, \textcolor{myc}{like multivalent ions, polyelectrolytes and polyampholytes whose adsorption-desorption mechanism represents one the key problem tackled in soft matter physics\cite{caetanoCriticalAdsorptionMultiple2020,caetanoCriticalAdsorptionPeriodic2017}}. \textcolor{myc}{Induced-demixing has been widely investigated in the past}. Domain and raft formation in membranes has been found in the presence of bivalent ions \cite{jacobsonPhaseTransitionsPhase1975}, RNA and DNA strands \cite{michanekRNADNAInteractions2010}, charged proteins \cite{mbamalaDomainFormationInduced2005}, and oppositely charged polyelectrolytes \cite{macdonaldPolyelectrolyteinducedDomainsLipid1998}. By contrast, induced mixing has been documented much less and only very recently it has been shown that the addition of functionalized nanoparticles \cite{canepaAmphiphilicGoldNanoparticles2020} or thermoplastic polymers can indeed give rise to a suppression of lipid segregation \cite{bochicchioInteractionHydrophobicPolymers2017}. The general mechanisms dictating a mixing induced by adsorbing molecules, polymers or colloids and the thermodynamic conditions under which this occurs, have not been explored neither understood in depth.  
Finally, domain formation and lipid miscibility plays a crucial role for the membrane permeability. Passive transport through biomembranes and bilayer lipid membranes is in many cases controlled by existing small holes (or voids) within them \cite{kotykMembraneTransport1977,kashchievBilayerLipidMembrane1983}.
There are several experimental evidences that proton, water and potentially drug permeability of mixed lipid membranes is enhanced by raft formation\cite{laroccaProvingLipidRafts2013,ghyselsPermeabilityMembranesLiquid2019,gensureLipidRaftComponents2006,sciollaInfluenceDrugLipid2021}, since inter-domain interfaces enhances ions and small molecules intrusion, and that membranes can be even destabilized by lipid substitution\cite{laroccaProvingLipidRafts2013} or multivalent ions\cite{haStabilizationDestabilizationCell2001}. In addition to biological membranes, hole formation have been used to describe the rupture of bilayer (Newtonian) black foam films \cite{derjaguinTheoryRuptureBlack,kashchievNucleationMechanismRupture1980}, the evaporation of liquids through adsorbed monolayers
\cite{dickinsonHardDiskFluid1978}, as well as the thermodynamic equilibrium of monolayers on liquid surfaces\cite{stoecklyEquationStateFattyacid1977}.\\

This said, despite an undisputed relevance of liquid mixtures lying on spheres and void formation in them, a systematic investigation through simple models on the role played by miscibility and its coupling to bulk particle adsorption in determining the mixture structure and in triggering void nucleation is still lacking.

In this work we introduce a minimal model to study the structure across a consolute point\cite{wildingLiquidvaporPhaseBehavior1998} of a symmetric Lennard-Jones binary mixture confined on a spherical surface by a harmonic potential, leaving particles free to oscillate along the radial direction. Such a quasi-2D model is employed first to characterize the structure of bare liquid mixtures at fixed temperature (here $T^*=0.5$) and varying miscibility parameter ($\aAB$) ruling the demixing of the two fluid components (A and B). We compute both the total and the partial geodesic pair correlation functions, and we introduce a geodesic mixing parameter ($\Xi$) through which we describe both the spontaneous demixing and the effect of adsorbing bulk particle (AP) addition on the mixture structure. We focused on the role of the AP preferential affinity with one of the two mixture components and we give a simple structural criterion to distinguish between AP induced mixing and demixing, showing that the former occurs only for nearly demixed systems with large spatial fluctuations of domain boundaries. 
For both AP-free mixtures and AP-decorated ones, we further investigate void size distributions. We show that in the presence of spatially fluctuating domains and inter-domain interfaces, large void formation is not favoured, and that AP adsorption can both enhance and disfavour the onset of large voids, depending on the miscibility of the two fluid constituents.        

\section{Model}
The binary mixture studied in this paper consists of
fluids ($A$ and $B$) made of spherical particles of the same
size, $\sigma_{AA}=\sigma_{BB}$, and at concentrations of 50\% each. Particles of the same type interact through a truncated 12–6 Lennard-Jones (LJ) potential\cite{diaz-herreraWettingPhenomenonLiquidvapor2004}:
\begin{equation}\label{LJ}
    \begin{cases}
U_{LJ}(r_{ij})=4\varepsilon_{ij}\left[\left(\frac{\sigma_{ij}}{r_{ij}}\right)^{12}-\left(\frac{\sigma_{ij}}{r_{ij}}\right)^{6}\right],&\mbox{if } r<3$\sAA$\\
0, &\mbox{if } r>3$\sAA$
\end{cases}
\end{equation}
with a mixing rule 
\begin{equation}
\sAB=\frac{1}{2}(\sAA+\sBB),\quad  \epsAB =\aAB\epsAA,
\end{equation}
where $\epsAA=\epsBB$, and $\aAB$ is the parameter controlling the miscibility of the two fluids.
Notice that when, $\aAB=0$, we obtain two independent LJ fluids
while in the opposite case, $\aAB=1$, the system reduces to a
single LJ fluid. By choosing, $0<\aAB<1$, the attractive part of
the interaction between dissimilar particles becomes weaker than that of the $AA$ and $BB$ interactions, favouring demixing\cite{diaz-herreraWettingPhenomenonLiquidvapor2004}. Each particle vibrates around the ideal surface of a sphere of radius $R_s$ since they are further confined via a parabolic potential: 
\begin{equation}
U_c(r_i)=\frac{1}{2}\kappa_c(r_{i}-R_s)^2,
\end{equation}
where $\frac{\kappa_c\sAA^2}{2k_BT}=10$ and $r_i$ is the distance between particle $i$ and the center of the confining sphere. To check the influence of the system size on demixing, we simulated three systems characterized by different particle numbers ($N_p=250,500,1000$) and the same area fraction occupied by the particles $\phi=\frac{N_p\sAA^2}{16R_s^2}=0.5$, corresponding to a reduced density $\rho^*=\frac{4\phi}{\pi}=0.63$. \textcolor{myc}{The choice of the system size has been made to ensure that the particle-to-sphere size ratio $\sAA/2R_s$, ranging here from 0.044 to 0.089, was comparable to the bilayer-to-vesicle size ratios encountered in small lipid vesicles \cite{sciollaInfluenceDrugLipid2021}, the latter being $\approx 0.05$.
The attractive interaction between the confined particles ($\leq 2k_BT$, see section \ref{simu}) and the energy scale ($\frac{\kappa_c\sAA^2}{2}$) of the confining potential have been chosen to reproduce the same order of magnitude respectively of the in-plane lipid-lipid interactions (few $k_BT$ units\cite{komuraLateralPhaseSeparation2004} for homogeneous systems and possibly less than $k_BT$ for dissimilar lipids\cite{almeidaThermodynamicsMembraneDomains2005}) and the bending energy of membranes per lipid (from few tens to hundreds of $k_BT$ per molecule \cite{pelitiBiologicallyInspiredPhysics1991}). The first allows molecules to diffuse and structure itself as a fluid and its variation determines whether demixing occurs at room temperature if more than one lipid type is present. The second gives rise to membrane stability. The magnitude of the confining potential has been further selected so as to preserve equipartition of kinetic energy during all the simulation runs and avoid flying ice-cube artifacts.} 

Free unconfined APs interacting with the mixture via the potential (\ref{LJ}) with LJ diameter $\sPP=\sAA$ have been further added to mixtures composed by $N_p=500$ particles in order to study the effect of physical adsorption on the structure of the confined AB-system. The interaction between APs and the mixtures is set by equation \ref{LJ}, where miscibility between APs and A-type particles has been fixed by setting their miscibility parameter to $\aAP=1$, while that between APs and B-type particles has been tuned by varying their mutual miscibility parameter (hereinafter $\aBP$) in the range $1-0.025$. In this way the interaction between the adsorbing particles and the confined mixtures is such as to be preferential (for $\aBP<\aAP$) or, let's say, symmetrical (for $\aBP=\aAP$) while the total adsorption energy decreases when $\aBP$ is lowered. 
\textcolor{myc}{In such a way the simulated systems mimic the weak physical adsorption of small molecules whose adsorption energies do not exceed few $k_BT$ units.}
The surface of the confining sphere has been further made impenetrable to bulk APs via the purely repulsive potential:
\begin{equation}\label{LJAPrep}
U_{LJ}(r_{\scriptscriptstyle PS})=4\varepsilon_{\scriptscriptstyle PS}\left[\left(\frac{2R_s+\sPP}{r_{\scriptscriptstyle PS}}\right)^{12}\right],
\end{equation}
where $\varepsilon_{\scriptscriptstyle PS}/(k_BT)=2\cdot 10^{-4}$ and $r_{PS}$ is the center-to-center distance between an AP and the sphere confining the AB mixture. 
\section{Simulation details}\label{simu}
We have carried out extensive MD simulations to investigate the structural properties of this model binary mixture as a function of $\aAB$ and $\aBP$. The equations of motion were integrated using a velocity-Verlet algorithm with reduced time step $\Delta t^*=\frac{\Delta t}{\sAA}\sqrt{\frac{\epsAA}{m_A}}=5\cdot10^{-6}$. $m_A$ is the particle mass for species $A$ and $m_A=m_B=m_P$. We fixed in all simulations the reduced temperature at $T^*=\frac{k_BT}{\varepsilon_{AA}}=0.5$, where 2D Lennard-Jones fluids are in a condensed liquid state \cite{barkerPhaseDiagramTwodimensional1981,holianFragmentationMolecularDynamics1988}.  
For the chosen set of miscibility parameters ($\aAB$,$\aBP$) we equilibrated the systems using the Berendsen’s thermostat with a large coupling constant $\tau_T=2\cdot 10^3\Delta t$ to obtain a stable trajectory in equilibrium. Equilibration runs have been performed for $2\cdot10^8$ time steps and production runs have been carried out in the NVE ensemble for \textcolor{myc}{$5.2$} $\cdot 10^7$ time steps during which total and kinetic energy drifts were absent. To exclude the presence of important ballistic collective motions of the confined particles due to an eventual rigid rotation of mixtures on the sphere, and energy transfer from high to low frequency modes, we have further performed long NVT simulations for $N_p=1000$ and $N_p=500$ at $\aAB=1$ to compute the mean angular square displacements: a pure diffusive behavior has been observed (see Appendix \ref{appA}).
To minimize correlations between measurements we calculated structural
quantities every $\Delta t_s=5\cdot 10^{\textcolor{myc}{5}}$ time steps (\textcolor{myc}{see Appendix \ref{appD}}). 
Simulations including $N_{\scriptscriptstyle AP}=1000$ free bulk particles have been carried out only for mixtures with $N_p=500$. The choice of such an excess of bulk particles has been inspired by the experimental system recently investigated by the author and coworkers \cite{sciollaInfluenceDrugLipid2021} in which lipid phase separation in mixed liposomes has been observed in excess of Isoniazide, one of the primary drugs used in the tuberculosis treatment. 
An investigation on the effect of bulk particle concentration, temperature and mixture composition are currently in progress and will be the subject of a future publication. Bare mixtures confined on the sphere have been simulated in a cubic box of size $L=10R_s$ with no extra periodic boundary conditions, with the sphere being at the center of the box. We underline here that the main characteristic of such systems is to be finite but without boundaries, with periodic boundary conditions being not an artifact of the simulations,
but an essential feature of it. Hence, in this case there is no need for further
assumptions or corrections: the complete system is simulated.
Conversely, we carried out simulations including bulk APs in a box of size $L=4.2R_s$, correspondent to a reduced density for the bulk particles $\rho^*=\frac{N_{\scriptscriptstyle AP}\sigma_{CC}^3}{(L^3-4/3\pi R_s^3)}=0.029$, with cubic periodic boundary conditions applied only to free APs.      

\section{Bare mixtures}
\begin{figure}[htbp]
\includegraphics[width=8cm]{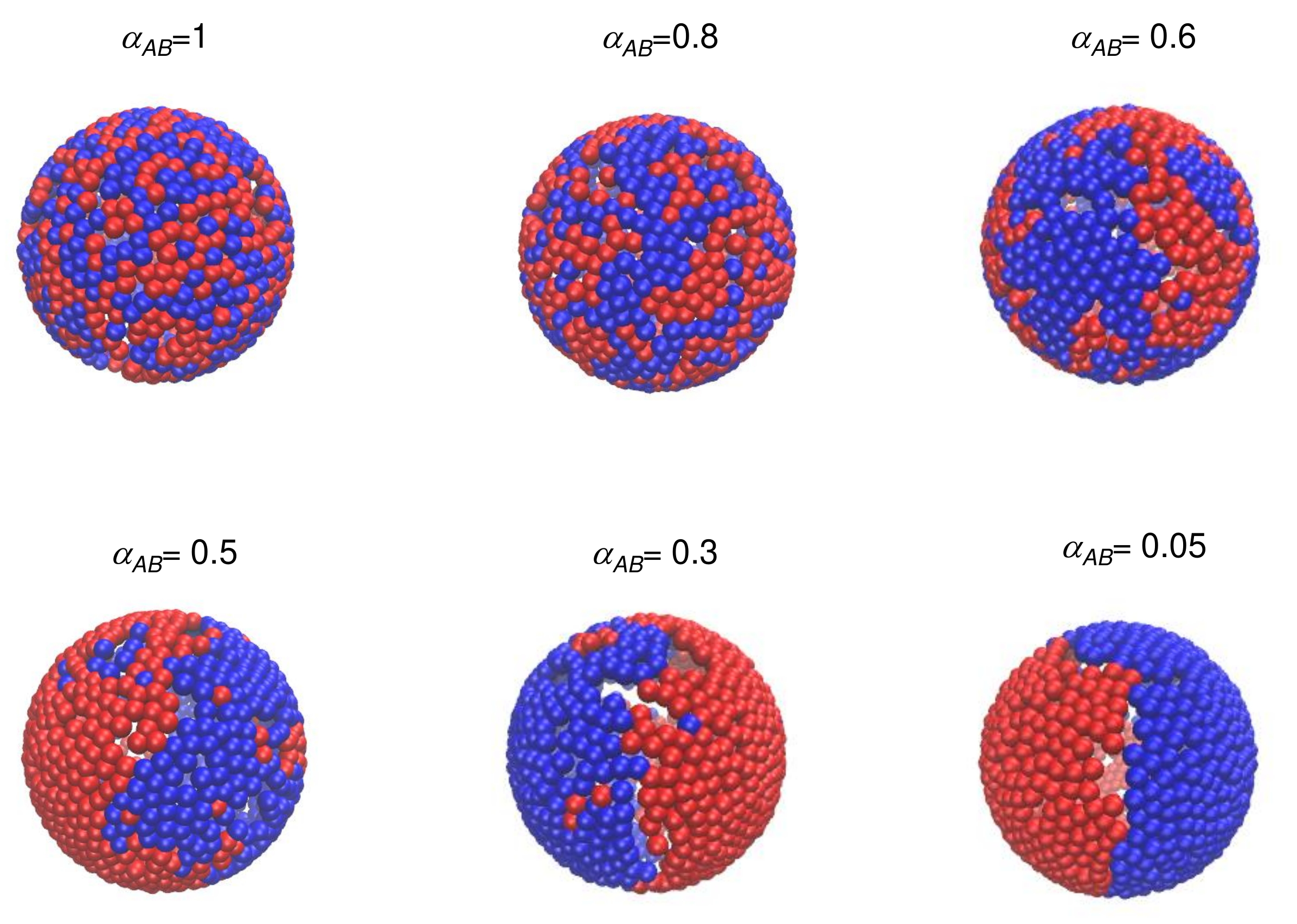}
\caption{Snapshots of binary 50:50 AB-mixtures ($N_p=1000$) confined to move on a spherical surface with radius of curvature $R_s=11.18\sAA$ and different miscibility parameters $0.05\leq\aAB\leq 1$, as indicated in the figure. All snapshots refer to equilibrium states.}\label{snapshots}
\end{figure}

The structure of bare mixtures has been studied by inspecting the total $g(s)$ and the partial $g_m(s)$ pair correlation functions defined as:
\begin{equation}\label{g(s)}
g(s)=\frac{1}{2N_p^2sin(s)} \frac{dn(s)}{ds}
\end{equation}
\begin{multline}\label{gAB(s)}
g_m(s)=\frac{1}{2}(g_{\scriptscriptstyle AB}(s)+g_{\scriptscriptstyle BA}(s))=\\ \frac{1}{2}\left[\frac{1}{2N_A^2sin(s)}\frac{dn_{\scriptscriptstyle AB}(s)}{ds}+\right.\\
\left.\frac{1}{2N_B^2sin(s)}\frac{dn_{\scriptscriptstyle BA}(s)}{ds}\right]
\end{multline}
where $s\in[0,\pi]$ is the normalized geodesic distance on the unit sphere, $dn(s)$ is the average number of particles at geodesic distance comprised between $s$ and $s+ds$ from one test particle and $dn_{\scriptscriptstyle AB}(s)$ ($dn_{\scriptscriptstyle BA}(s)$) is the average number of particles of type B (A) at geodesic distance comprised between $s$ and $s+ds$ from a test particle of type A (B). For homogeneous mixtures ($\epsAA=\epsBB=\epsAB$) $g_m(s)=g(s)$, while the largest difference between the two functions is attained when the two fluids are fully separated on the sphere in a Janus-like configuration.\\
Figure \ref{snapshots} shows snapshots of equilibrated samples of mixtures for $N_p=1000$ with different mixing parameters $\aAB$ ranging from 1 down to 0.05. A continuous transition from fully miscible to Janus-like patterns characterizes the demixing transition with an intermediate regime $0.5\leq\aAB\leq 0.8$ where large spatial interfacial fluctuations and domains emerge.

\begin{figure}[htbp]
\includegraphics[width=8cm]{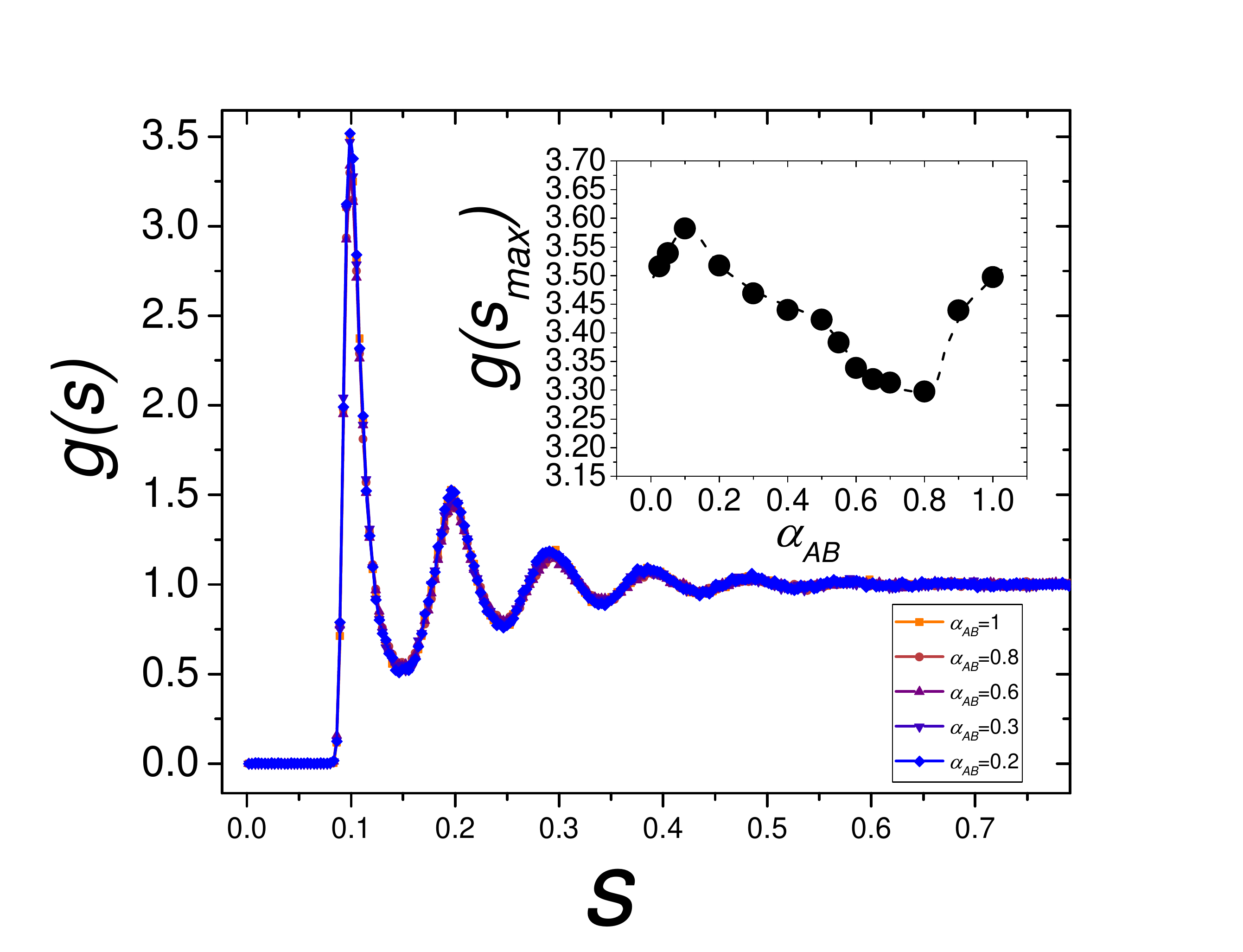}
\caption{Total pair correlation functions $g(s)$ for $N_p=1000$ and selected values of the miscibility parameter $0.2\leq\aAB\leq1$. Inset: height of the first peak $g(s_{max})$ for the whole set of miscibility parameters investigated in this work.}\label{grtot-noions}
\end{figure}

\begin{figure}[htbp]
\includegraphics[width=8cm]{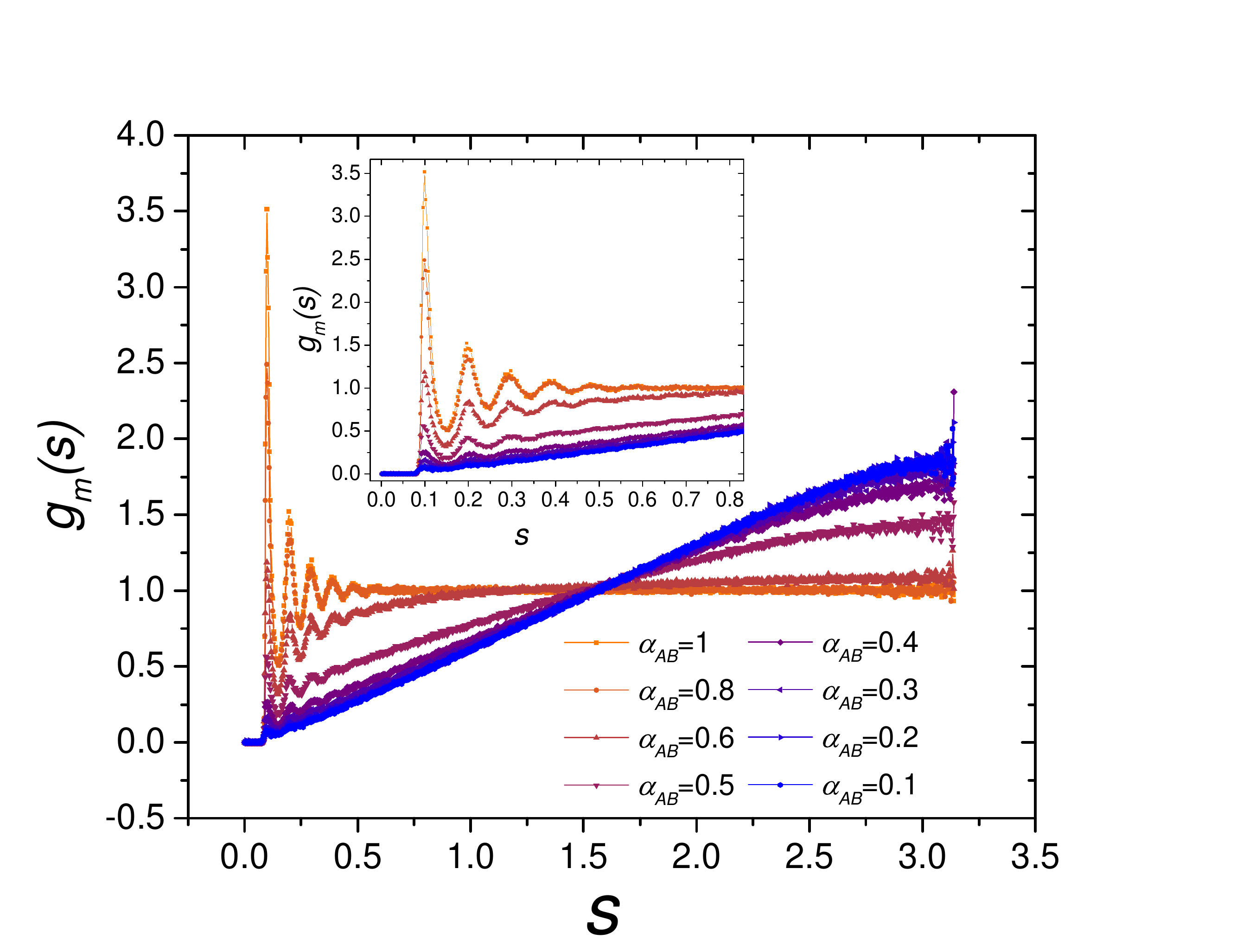}
\caption{Partial pair correlation functions $g_m(s)$ for $N_p=1000$ and selected values of the miscibility parameter $0.1\leq\aAB\leq 1$. Inset: Same data of the main panel for $0\leq s\leq 0.8$.}\label{grAB-noions}
\end{figure}

By computing both $g(s)$ (Figure \ref{grtot-noions}) and $g_m(s)$ (Figure \ref{grAB-noions}) we follow both the structure of the whole fluid and the demixing of the two components on the sphere. Though demixing does not affect remarkably the local structure, we unambiguously observe a weak melting of the mixture induced by the formation of large domains, as signalled by the global minimum of the height of the first peak of $g(s)$ as a function of $\aAB$. The decrease of the first peak reveals that the average local density of the fluid in the immediate vicinity of any particle decreases with respect to the homogeneous system when fluctuating domains coexist (inset of Figure \ref{grtot-noions}), since a complete structural demixing is not achieved, the number of weak AB bonds is still large and particles are on average less bound. On the contrary, when the Janus-like configuration is attained and small spatial fluctuations characterize the interface between the A-rich and the B-rich phases, the number of AB contacts decreases and the average particle-particle distance in the fluid goes back to about the one obtained for $\aAB=1$.       

\begin{figure}[htbp]
\includegraphics[width=8cm]{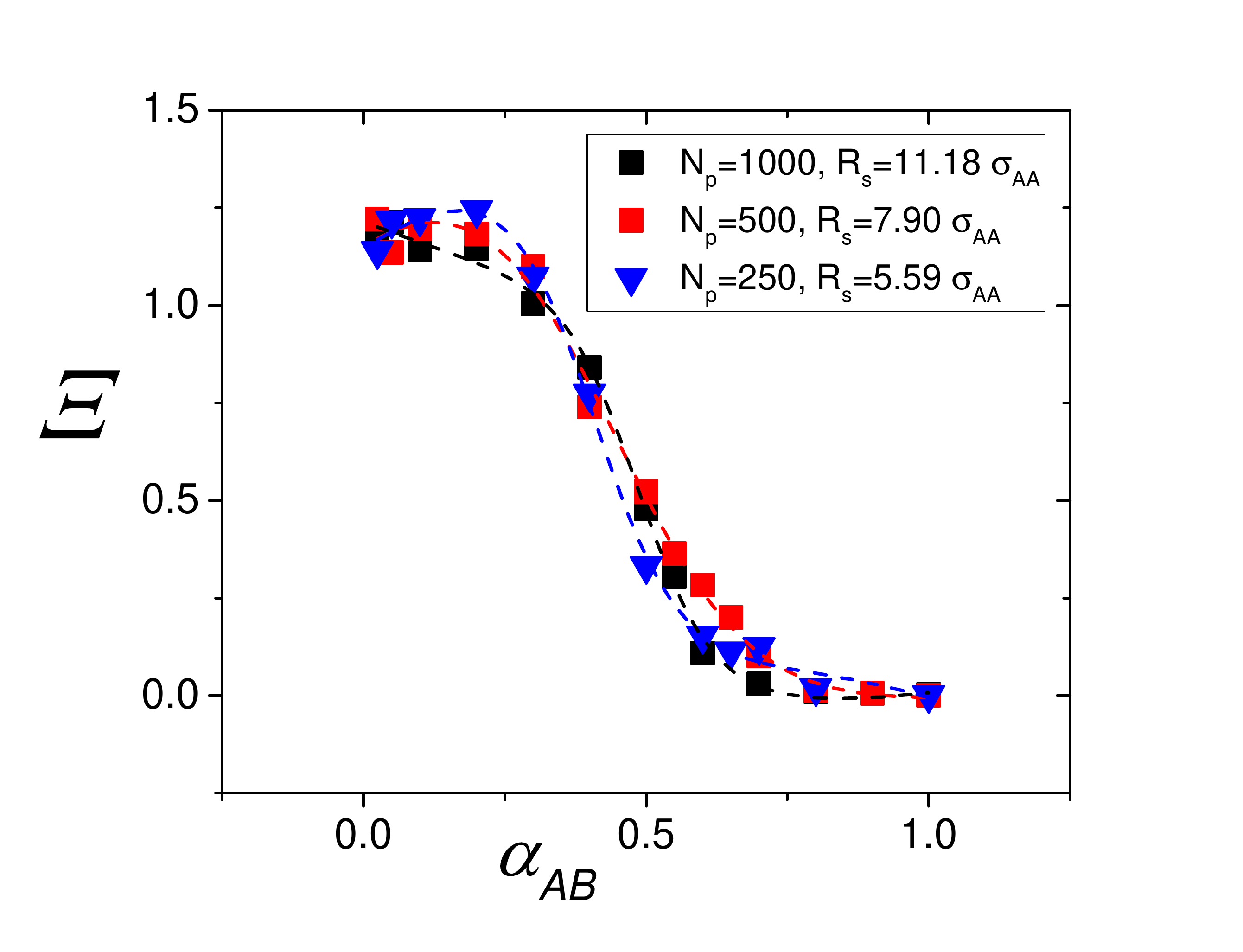}
\caption{Geodesic mixing parameter $\Xi$ \textcolor{myc}{as a function of} the miscibility parameters $\aAB$ for the three different system sizes $N_p=250,500,1000$, and fixed area fraction \textcolor{myc}{$\phi=0.5$}. Dashed line are fits obtained via equation \ref{fitxi} (Appendix \ref{appB}).}\label{distmix-energy}
\end{figure}

\begin{figure}[htbp]
\includegraphics[width=8cm]{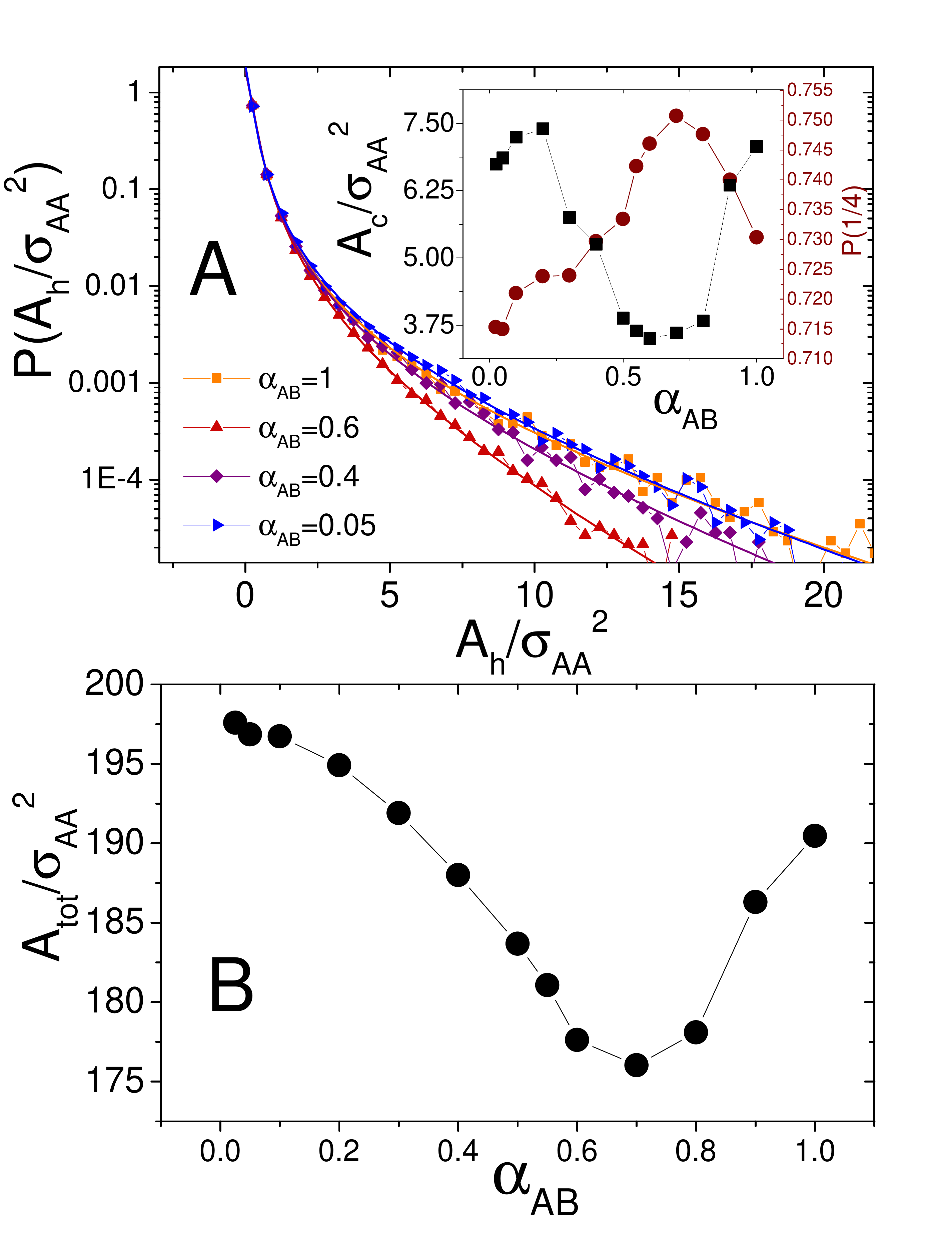}
\caption{Panel A: Void size distribution $P(A_h/\sigma_{AA}^2)$ for different miscibility parameter $\aAB$ as indicated in the panel. The inset shows the cut-of area $A_c$ (black sqaures) defined in equation \ref{Pvoid} and the fraction the smallest voids $P(1/4)$ (blue circles) as a function of $\aAB$. Panel B: Total void area in function of $\aAB$. The area takes into account only voids with size $A_h\geq\sAA^2/4$.}\label{voids-noions}
\end{figure}

The partial correlation functions $g_m(s)$ (Figure \ref{grAB-noions}) follow the demixing of the two components: when particles are fully mixed $g_m(s)=g(s)$, while when $\aAB\ll 1$ and full demixing occurs, the probability to find an AB contact drastically decreases (small $s$ region) and it increases progressively until it reaches its maximum at angular distance $s=\pi$. This is accompanied by the disappearence of the correlation peaks of $g_m(s)$ when $\aAB$ decreases, as shown in the inset of Figure \ref{grAB-noions} for representative values of $\aAB$.\\
Since confining spherical surfaces are a closed manifold with uniform curvature, we can describe in a nutshell the demixing by defining a geodesic mixing parameter as the square distance between $g(s)$ and $g_m(s)$ on $[0,\pi]$:
\begin{equation}\label{xi}
\Xi=\|g(s)-g_m(s)\|^2=\int_0^{\pi}[g(s)-g_m(s)]^2ds.
\end{equation}
$\Xi$ is zero when the two components of the mixtures are fully mixed and larger than zero to an extent depending on the structural demixing.   
In Figure \ref{distmix-energy} we show $\Xi(\aAB)$ obtained for the three simulated system sizes ($N_p=250,500,1000$). $\Xi(\aAB)$ smoothly passes from very low values compatible with zero (given the simulation noise) for $\aAB=1$ to larger values, that depends weakly on $\aAB$ when the latter is lower than $\approx 0.2$ for each system size, pointing out that radius of curvature $R_s$ does not impact remarkably the structural demixing transition apart from a very small shift of the critical miscibility parameter $\aAB^c$ defined as the point where $\frac{d^2\Xi}{d\aAB^2}=0$, that we extract as detailed in Appendix \ref{appB}.
Actually, we expect that compositional fluctuations start to play a role for very small radius of curvatures impacting more the angular particle distribution, favouring mixing.
To gain a complementary insight on the structure of the mixtures and their stability against void formation we computed the normalized large-void size distribution $P(A_h/\sAA^2)$ defined as 
\begin{equation}\label{def-Pvoid}
P(A_h/\sAA^2)=\frac{N(A_h/\sAA^2)}{\sum\limits_{h=1}^{\infty}N(A_h/\sAA^2)},\quad \mbox{with\hspace{0.2cm}} A_h\geq \frac{\sAA^2}{4}
\end{equation}
where $A_h$ is the area of a void with size $h\sAA^2/4$ and $N(A_h/\sAA^2)$ is the number of voids with normalized size $A_h/\sAA^2$. We therefore restrict the search of voids to those having area larger than a particle radius square, since we are not interested to the distribution of small interstitial spaces between particles but rather to those voids that are compatible with inclusion of adsorbing particles or molecules in real systems. Indeed the minimal void-to-particle size ratio taken into consideration in this work ($A_h/\sAA^2=1/4$) is of the same order of magnitude of that characterizing water molecules and lipid cross section in membrane lipids\cite{israelachviliModelPackingLipids1975}.     
To compute each void area $A_h$ we applied a spherical grid to the confining sphere, where $N_{g}$ grid points are almost-equidistant and have been generated following the scheme proposed  in\cite{desernoHowGenerateEquidistributed}, with a lattice spacing $a \simeq \sAA/2$.
We consider as a void a subset of $n_g$ grid points such that: i) each point of the subset is far from any particles by a geodesic distance larger than $3/4\sAA$; ii) a void has $n_g>1$ if each point of the subset have at least one grid point distant one lattice spacing and belonging to the same subset (cluster of points). The area of the void is then well approximated by $A_h=n_g \sAA^2/4$, with a minimum measurable value equal to $\sAA^2/4$. In figure \ref{voids-noions} we report the normalized distribution of void sizes for selected values of $\aAB$. The reentrant behavior of the distribution tails goes along with the already mentioned fluidization of the system (Figure \ref{grtot-noions}) when jagged domain boundaries characterize the mixtures. In this regime, since $AB$ contacts are maximized and particles are on average less bound, the formation of large voids is hampered. To describe exhaustively this behavior we have fitted the void distributions with a phenomenological function of the following type:
\begin{equation}\label{Pvoid}
P(A_h/\sAA^2)=P_0+\frac{a}{1+\left(b\frac{A_h}{\sAA^2}\right)^c}\cdot e^{-A_h/A_c} 
\end{equation}
where $A_c$ is the exponential characteristic cutoff area of the distribution, $P_0$ is an offset set to zero for distribution showing exponential tails and larger than zero when large voids are created by bulk particle adsorption (see section \ref{withads}). Finally $a$, $b$ and $c$ characterize the power law regime observed for small void sizes. 
The choice of the fitting function has not been made by chance. Boutreux and  De Gennes \cite{boutreuxCompactionGranularMixtures1997} first postulated exponential decays for void size distributions in granular matter and based on this, Caglioti and coworkers \cite{cagliotiCooperativeLengthApproach1999} related quantitatively the characteristic dynamical properties for the same systems to quasistatic quantities, e.g. free-volume and configurational entropy. In our mixtures, however thermal (and stress) fluctuations are important and the void formation should also be related to the probability of spontaneous rupture events of a given size. Rupture events often show power-law size distributions \cite{amitranoVariabilityPowerlawDistributions2012}
with an exponential cut-off due to the finite size of the system, while shifted gamma functions with exponential tails have proven to be appropriate to describe the statistics of void sizes in packed protein cores and jammed packings of amino-acid-shaped particles \cite{treadoVoidDistributionsReveal2019} and attractive emulsion droplets\cite{jorjadzeAttractiveEmulsionDroplets2011}. These Boltzmann-type distribution tails are those maximizing configurational entropy in packed systems \cite{jorjadzeAttractiveEmulsionDroplets2011}.
Our results are in line with these previous findings and point to void distribution tails with two distinct regimes: a power-law decay followed by an exponential cut-off. Finally, the deviation from a simple power-law at very low void areas ($\lim_{A_h\to 0}P(A_h/\sAA^2)=P_0+a$) reflects the fact that the void size is limited by the interstitial space between LJ sphere, so that the distribution must converge to 0 as $A_h\rightarrow 0$\cite{treadoVoidDistributionsReveal2019}. Such a deviation is found systematically for all the sets of parameters employed in our simulations, hence the need to use equation \ref{Pvoid} to fit satisfactorily all our data. A finite non-zero value for $P_0$ further points to a non-exponential (fat) tail of the distribution. This will be the case of some of the mixtures in presence of APs.       
The inset of Figure \ref{voids-noions}-A shows $A_c/\sAA^2$ together with the probability to have unit voids ($A_h/\sAA^2=1/4$). The two quantities show two non-monotonic trends with opposite concavity, corroborating the scenario in which demixing on small confining spheres is structurally a smooth process with the formation of small (large) voids being enhanced (suppressed) in the presence of fluctuating domains.
Similarly to what we observed for the cut-off areas, the total (large-)void size $A_{tot}=\sum\limits_{h=1}^{\infty}N(A_h/\sigma_{AA}^2)$shows a minimum for mixtures with large fluctuating domain interfaces (Figure \ref{voids-noions}-B). This further suggests that the integrity of spherical mixtures is stabilized by compositional fluctuations.    
All in all, the driving forces dictating large void formation are two: the ability of particles to form long-lived clusters, that is enhanced by increasing the total cohesive energy (here increasing $\aAB$) and dewetting, occurring for fully demixed fluids ($\aAB\ll 1$) where void formation is favored in the inter-domain regions. 

\section{Mixtures with adsorbing particles}\label{withads}
\begin{figure*}[htbp]
\includegraphics[width=\textwidth]{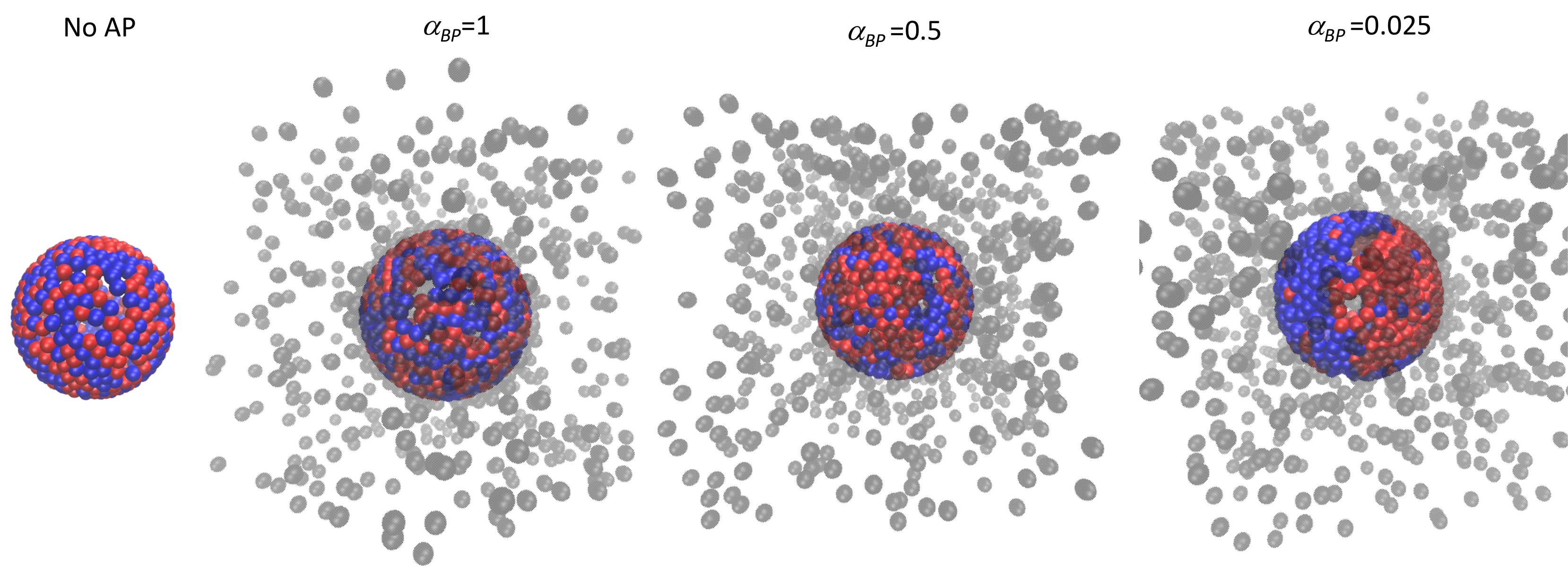}
\caption{Snapshots of binary 50:50 AB-mixtures ($N_p=500$) confined to move on a spherical surface with radius of curvature $R_s=7.90\sAA$, with $\aAB=1.0$ and 3 selected AP-mixture interaction parameter $\aBP$. A snapshot taken in absence of AP is given for reference. All snapshots refer to equilibrium states. By decreasing the affinity between APs and B-particles (snapshots from left to right), particle adsorption favours particle demixing and domain formation on the sphere, until reaching a Janus-like configuration for $\aBP=0.025$.}\label{Snap1}
\end{figure*}

\begin{figure*}[htbp]
\includegraphics[width=\textwidth]{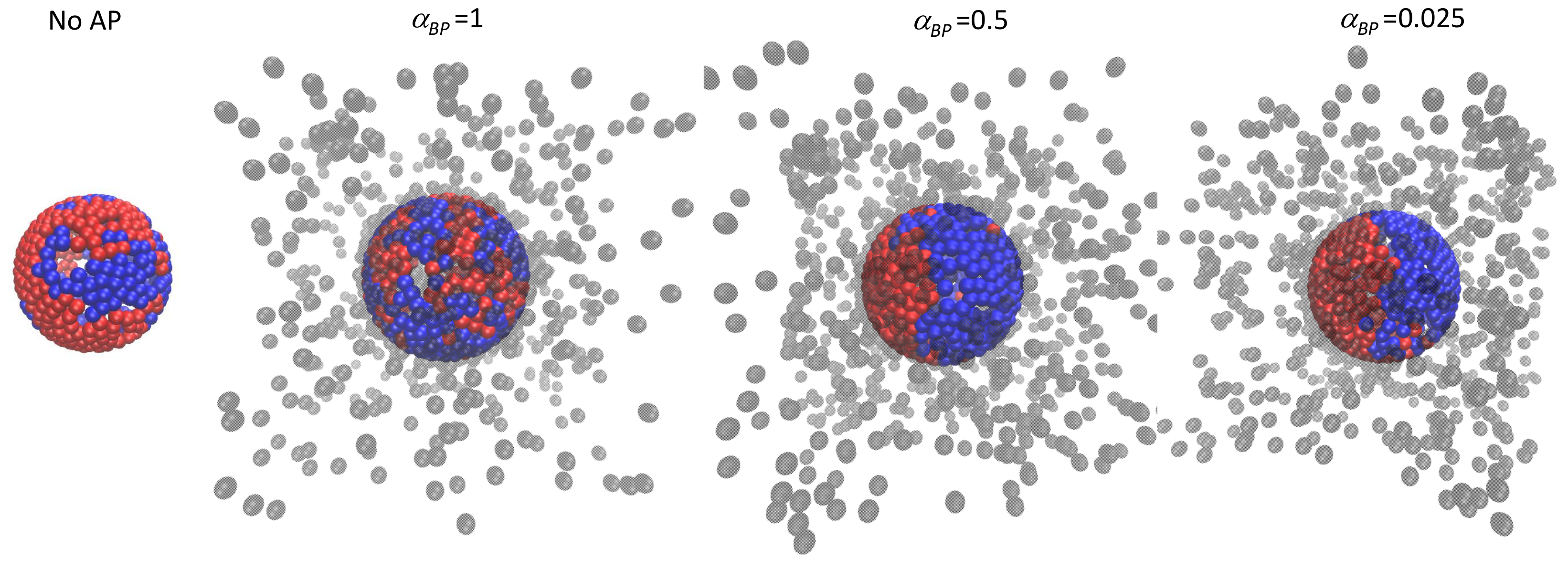}
\caption{Snapshots of binary 50:50 AB-mixtures ($N_p=500$) confined to move on a spherical surface with radius of curvature $R_s=7.90\sAA$, with $\aAB=0.6$ and 3 selected AP-mixture interaction parameter $\aBP$. A snapshot taken in absence of AP is given for reference. All snapshots refer to equilibrium states. Particle adsorption favours particle mixing first ($\aBP=1$) and then, by decreasing the affinity between APs and B-particles (snapshots from left to right), domain formation is progressively promoted on the sphere until reaching a Janus-like configuration ($\aBP=0.025$)}\label{Snap06}
\end{figure*}

\begin{figure*}[htbp]
\includegraphics[width=\textwidth]{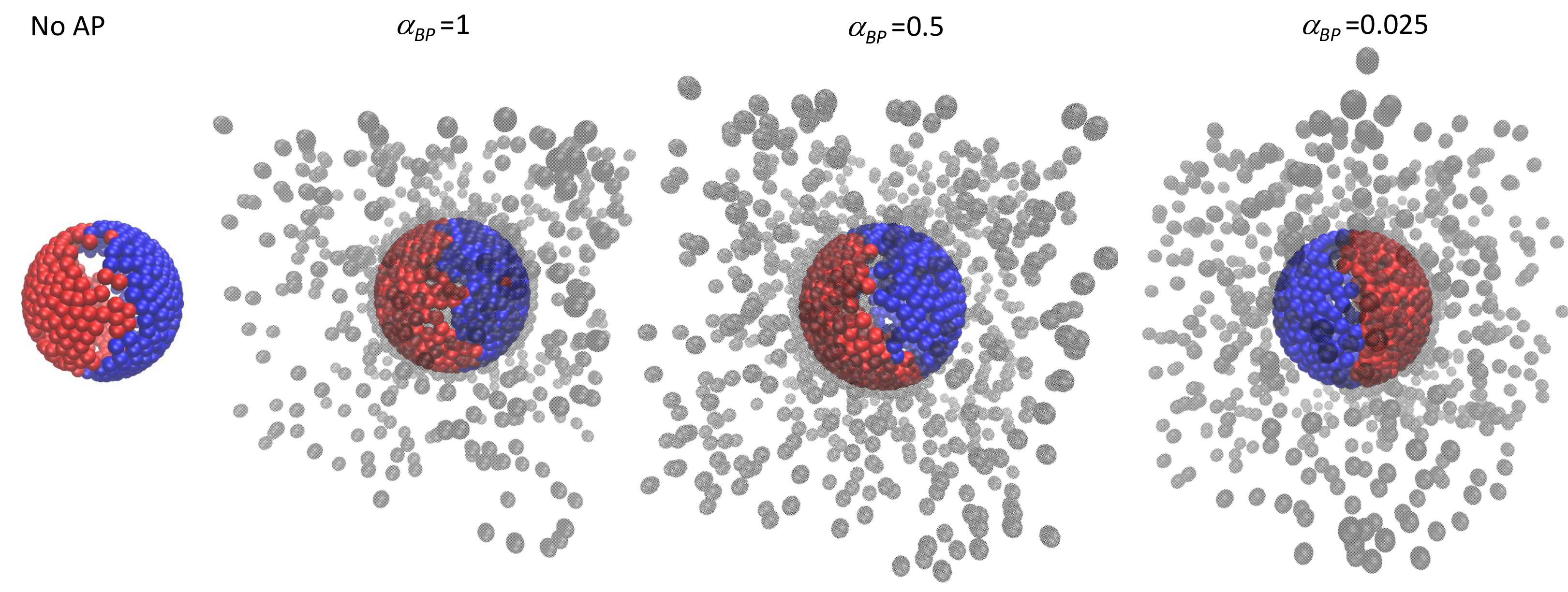}
\caption{Snapshots of binary 50:50 AB-mixtures ($N_p=500$) confined to move on a spherical surface with radius of curvature $R_s=7.90\sAA$, with $\aAB=0.05$ and 3 selected AP-mixture interaction parameter $\aBP$. A snapshot taken in absence of AP is given for reference. All snapshots refer to equilibrium states. In this case particle adsorption does not relevantly impact the initially demixed state of the confined fluid.}\label{Snap005}
\end{figure*}

The adsorption of particles on the confined mixtures modifies their structure and, largely affect the mixing-demixing transition. Figures \ref{Snap1}, \ref{Snap06} and \ref{Snap005} show selected equilibrium snapshot with $N_{\scriptscriptstyle AP}=1000$ APs for three representative cases: fully mixed fluids ($\aAB=1$), fluctuating domains ($\aAB=0.6$) and fully demixed fluids ($\aAB=0.05$). and we show also, for comparison, the configurations in the absence of APs. For fully mixed fluids, the addition of bulk particles with decreasing $\aBP$ coefficients gives rise to a progressive induced-demixing on the sphere. This is due to the large energy loss attained when bulk particles adsorb only on a condensed domain rich in one of the two species on the sphere that hence tend to segregate. In this case adsorbed particle can be viewed as energetic bridges between particles of type A.\\
In the intermediate regime (Figure \ref{Snap06}), where large spatial compositional fluctuations characterize the confined mixture, particle adsorption induces contrasting effects, depending on whether the interaction of APs with the particles belonging to the mixtures is symmetric ($\aBP=1$) or not. When $\aBP=1$ large domains break up, pushing the mixture towards mixing. Such a non-trivial behavior is the result of the subtle balance between the configurational entropy of the APs, that adsorb uniformly on the sphere when $\aBP=1$, lowering the line tension between the domains and causing their dissolution, and the energy gain that takes place when more energetic AA (or BB) bond are replaced by AB bonds.           
By contrast, when lower values of $\aBP$ are used and AP interact preferentially with particle of type A, the mixture adopts a Janus-like (fully separated) configuration, as already discussed for $\aAB=1$.\\Finally for low miscibility (Figure \ref{Snap005}) particle adsorption does not affect much the mixture structure and only a very weak effect can be observed when $\aBP$ is progressively lowered.
We quantify all these changes on the structure of the confined mixture by computing the total pair correlation function $g(s)$ and the partial correlation function $g_m(s)$ for the AB-mixtures, and the normalized structural mixing parameter
\begin{equation}\label{xinorm}
\Xi^*=\frac{\Xi_{\scriptscriptstyle AP}}{\Xi_{bare}},
\end{equation}
where $\Xi_{\scriptscriptstyle AP}$ and $\Xi_{bare}$ are calculated according to equation \ref{xi} in the presence and absence of adsorbing particles, respectively. 
The visual inspection of the mixture structure can be substantiated quantitatively in a first instance by comparing $g_m(s)$ with and without APs.
\begin{figure*}[htbp]
\includegraphics[width=\textwidth]{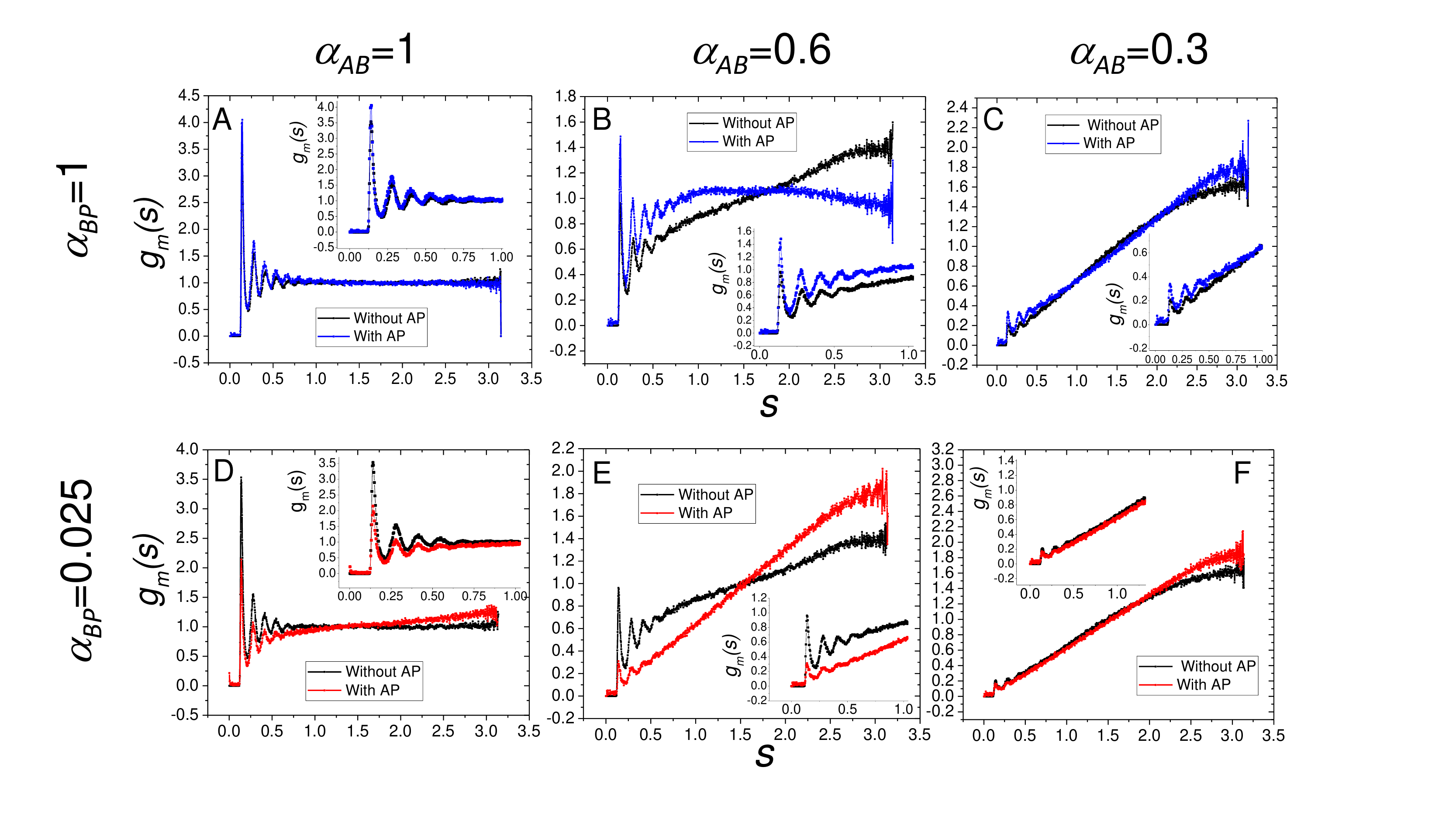}
\caption{Partial pair correlation functions $g_m(s)$ for symmetric AP-mixture interaction parameter $\aBP$=1 (panels A,B,C), large preferential affinity of the APs and A-type particles, $\aBP$=0.025 (panels D,E,F), and three selected miscibility parameters $\aAB$ as indicated in the figure. In all the cases $g_m(s)$ of the mixtures in absence of APs is also reported for reference. Insets show the same data in the range $0\leq s\leq 1$ to zoom on the AP effect on the local structure.} \label{grAB-ionseffect}
\end{figure*}
This is shown in figure \ref{grAB-ionseffect} for $\aAB=1$ (panels A,D), $\aAB=0.6$ (panels B,E) and $\aAB=0.3$ (panels C,F). We show the two representative cases of fully symmetric interactions between the APs and two components of the mixture ($\aBP=1$, panels A,B,C), and the most asymmetric case investigated ($\aBP=0.025$, panels D,E,F). The remarkable effect of APs on the mixtures at $\aAB=0.6$ catches immediately the eye, with contrasting relative variations passing from symmetric interactions (Figure \ref{grAB-ionseffect}-B) to strongly asymmetric ones (Figure \ref{grAB-ionseffect}-E). This is not observed for the fully miscible case ($\aAB=1$) or for Janus-like mixtures ($\aAB=0.3$), for which a decrease in ($\aBP$) gives rise to a progressive induced-demixing, the extent of the latter being much more pronounced for $\aAB=1$. In more detail, when large domain fluctuation dominates and AP addition induces particle mixing partial correlation functions show augmented local structure peaks and lower values (closer to one) at large distance with respect to the bare case. This is the case $\aAB=0.6$ and $\aBP=1$ (Figure \ref{grAB-ionseffect}-B). On the other hand when AP addition gives rise to induced demixing, $g_m(s)$ show reduced correlation peaks at small geodesic distance while it increases at large distance with respect to the bare system. This is the case $\aAB=0.6$ and $\aBP=0.025$ (Figure \ref{grAB-ionseffect}-E). For the other values of the miscibility parameter, namely $\aAB=1$ and $\aAB=0.3$, the AP addition has a relatively weak effect on the structure of the mixtures when $\aBP=1$ (Figure \ref{grAB-ionseffect}-A,C), while it produces an important fluid demixing of the initially miscible system ($\aAB=1$) for strongly asymmetric AP-mixture interactions ($\aBP=0.025$, Figure \ref{grAB-ionseffect}-D). This, again, is reflected by the net decrease of the structure peaks at small $s$ and a continuous increase of $g_m(s)$ at large $s$.        
\begin{figure}[htbp]
\includegraphics[width=8.0cm]{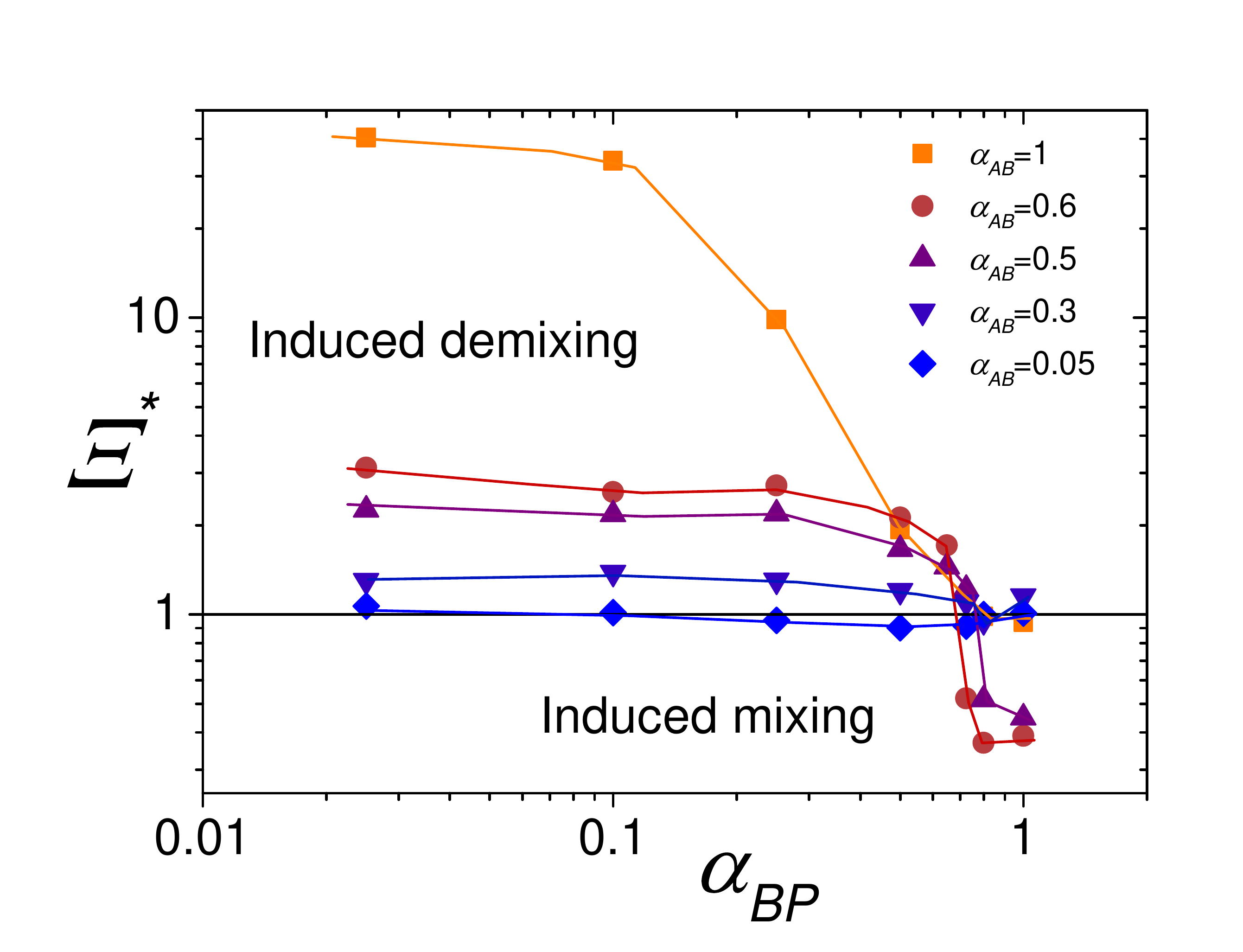}
\caption{Normalized geodesic mixing parameter $\Xi^*$ in function of the interaction parameter $\aBP$ for the $\aAB$ values indicated in the figure. Solid lines are only a guide to the eye.}\label{normonorm}
\end{figure}
To build up and visualize more clearly a general scenario based on all our results we computed $\Xi^*(\aBP)$ for all the investigated systems in the presence of AP. This is shown in figure \ref{normonorm} for $\aAB=1,0.8,0.5,0.3,0.05$. We note that: i) for large asymmetries of the AP-mixture interaction ($\aBP\ll 1$) the extent of the induced demixing increases for increasing values of $\aAB$; ii) the dependence of $\Xi^*$ on $\aBP$ weakens when AB miscibility decreases, reflecting the fact that the structure is affected by AP addition if particle segregation is poor;
iii) mixtures that are characterized by large fluctuating domains and interfaces between A-rich domains and B-rich domains show both induced demixing ($\Xi^*>1$) and induced mixing ($\Xi^*<1$) depending on the magnitude of the asymmetry $\aBP$ of the AP-mixture interactions. This is quite remarkable since many experimental systems are close to a demixing line and shows that both mixing and demixing are possible in such systems when particles or complex molecules, other than those composing the mixtures, are co-suspended.     
\begin{figure}[htbp]
\includegraphics[width=8.0cm]{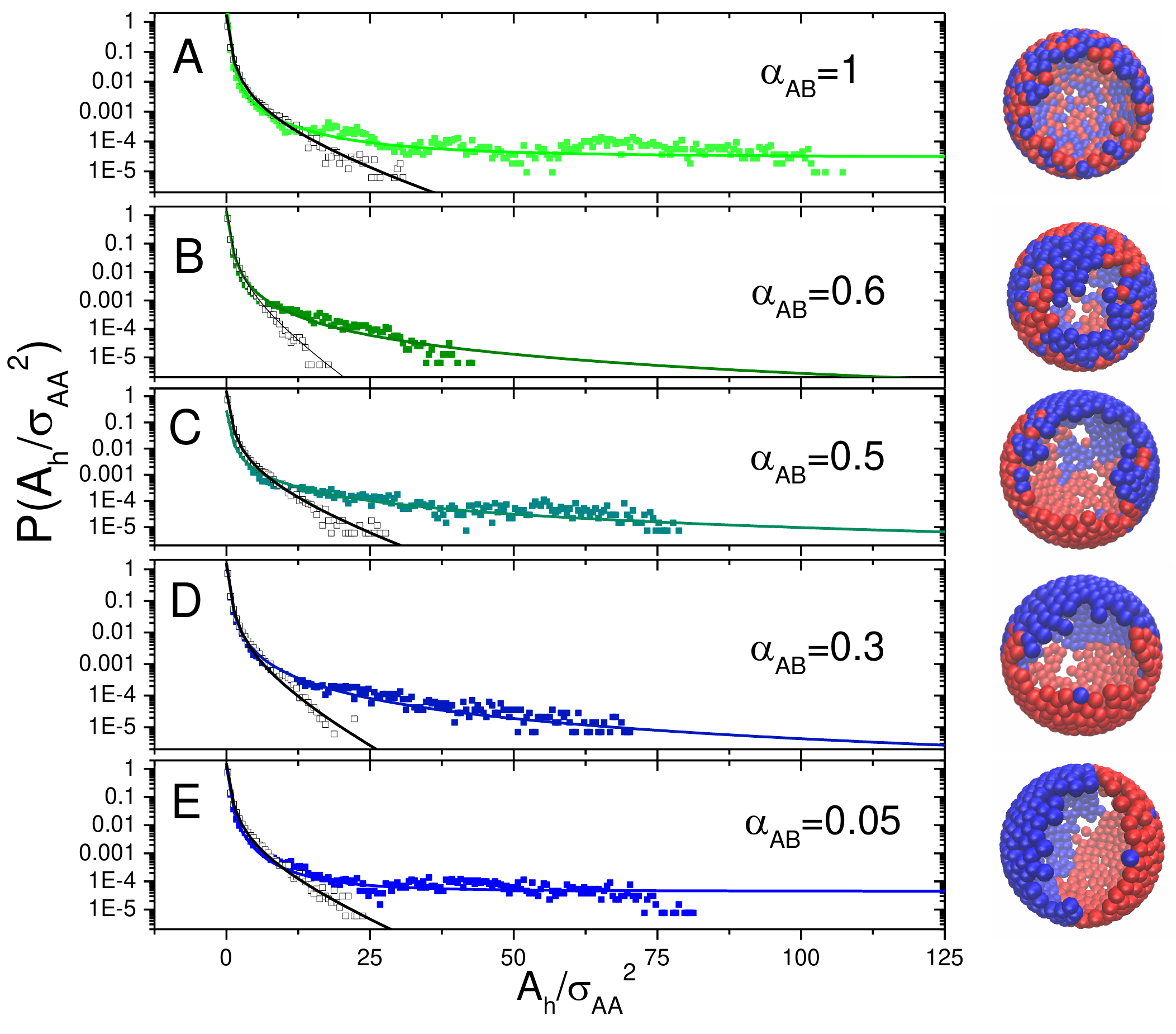}
\caption{Void size distribution $P(A_h/\sAA^2)$ with no preferential affinity ($\aBP=1$) and varying miscibility parameter $\aAB$ as indicated in the panels (A,B,C,D,E). Distributions obtained in absence of APs are also reported for reference. Solid lines are best fits obtained via equation \ref{Pvoid}.
For each $\aAB$ value one snapshot of an equilibrium configuration with $\aBP=1$ shows the biggest visible void. APs are not shown for clarity.}\label{voids-1-1}
\end{figure}
Finally, as done for bare mixtures, we report the void size distributions for representative systems in the presence of APs and we separate the case $\aBP=1$ from the rest of the systems since only in the former case we observe the formation of very large deviations from a simple exponential decay of the distribution tails. Figure \ref{voids-1-1} (panels A,B,C,D,E) shows the distributions $P(A_h/\sAA^2)$ and five representative configuration snapshots obtained for $\aBP=1$ and $\aAB=1,0.6,0.5,0.3,0.05$. The distributions are further compared to those obtained for bare mixtures. In all the cases, the effect of the AP addition is to enhance the formation of large voids since particle adsorption increases the cohesive energy between the particles confined on the sphere. This results in the suppression of the exponential tails of the void size distributions and the onset of fat tails, that we attribute to the transition from a regime where voids are uniformly distributed on the sphere (no APs) to one where one very large void, whose size fluctuates, dominates. In all such cases we find tails compatible with a power-law decay ($A_c=\infty$) with $P_0>1$ for $\aAB=1$ and $\aAB=0.05$ (Figure \ref{voids-1-1}-A,E).
However, such enhancement of large void formation, though present for each $\aAB$, does depend on the mixture structure. First of all, a reduced distribution widening is observed for intermediate $\aAB$, where large compositional fluctuations characterize the mixtures (see the $\aAB=0.6$ case in figure \ref{voids-1-1}-B). This is due to the large free energy cost that a large void would produce, since it would suppress domain fluctuations and enhance segregation while, in such a regime, the bare mixture is on average less packed (see Figure \ref{grtot-noions}) and irregular interfaces between domains form. Thus, the appearance of many voids of intermediate size is favoured at the expense of one very large void with fluctuating size. 
Upon decreasing further the value of $\aAB$ large voids develop again and, for Janus-like configurations, are spatially located at the interface between the A-rich domain and the B-rich domain, since the weaker bonds are those between dissimilar particles. Such interfacial void formation is reflected again by the extended (non-exponential) tails of the distributions $P(A_h/\sAA^2)$.\\
\begin{figure}[htbp]
\includegraphics[width=8.0cm]{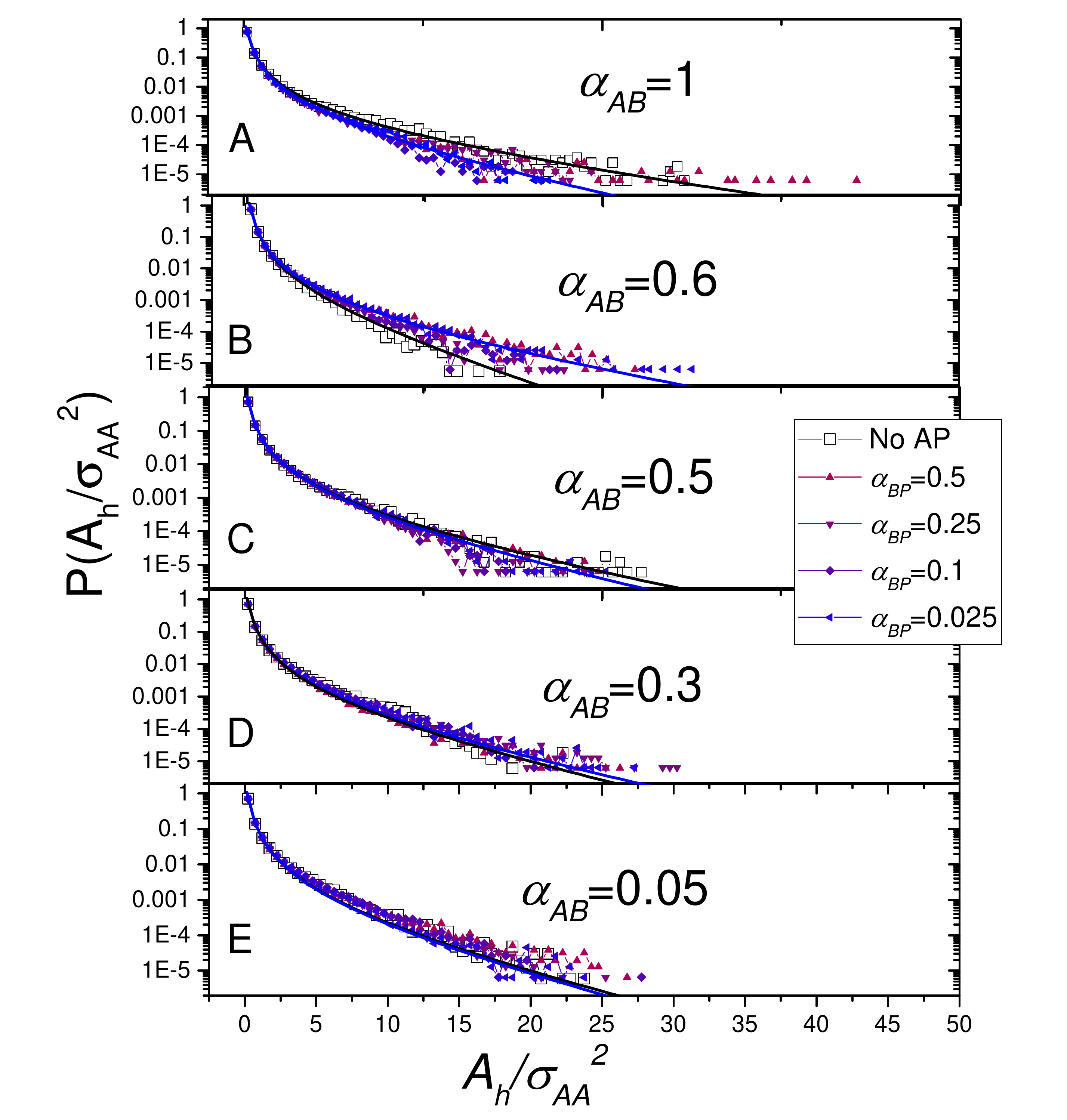}\\
\includegraphics[width=4.0cm]{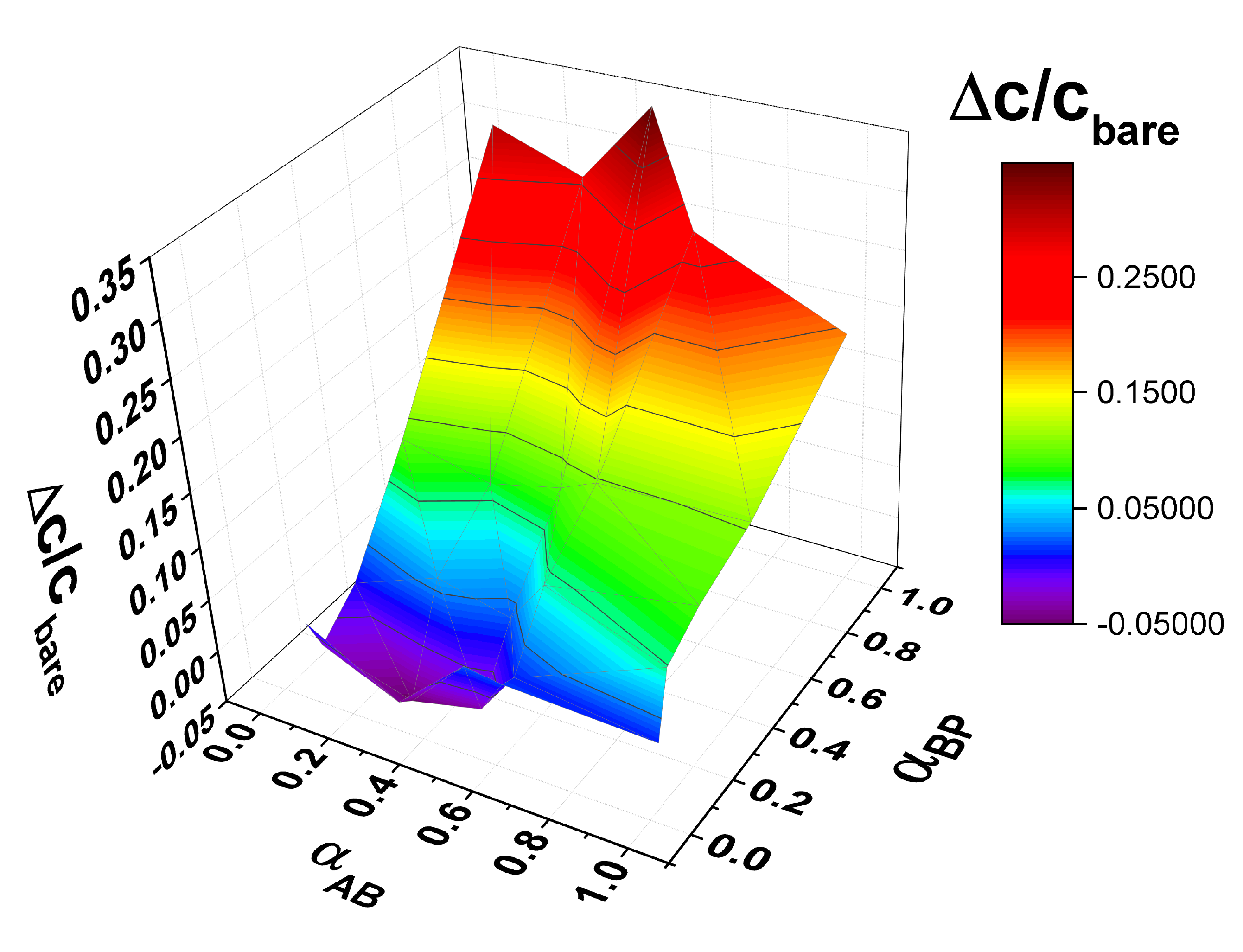}\includegraphics[width=4.0cm]{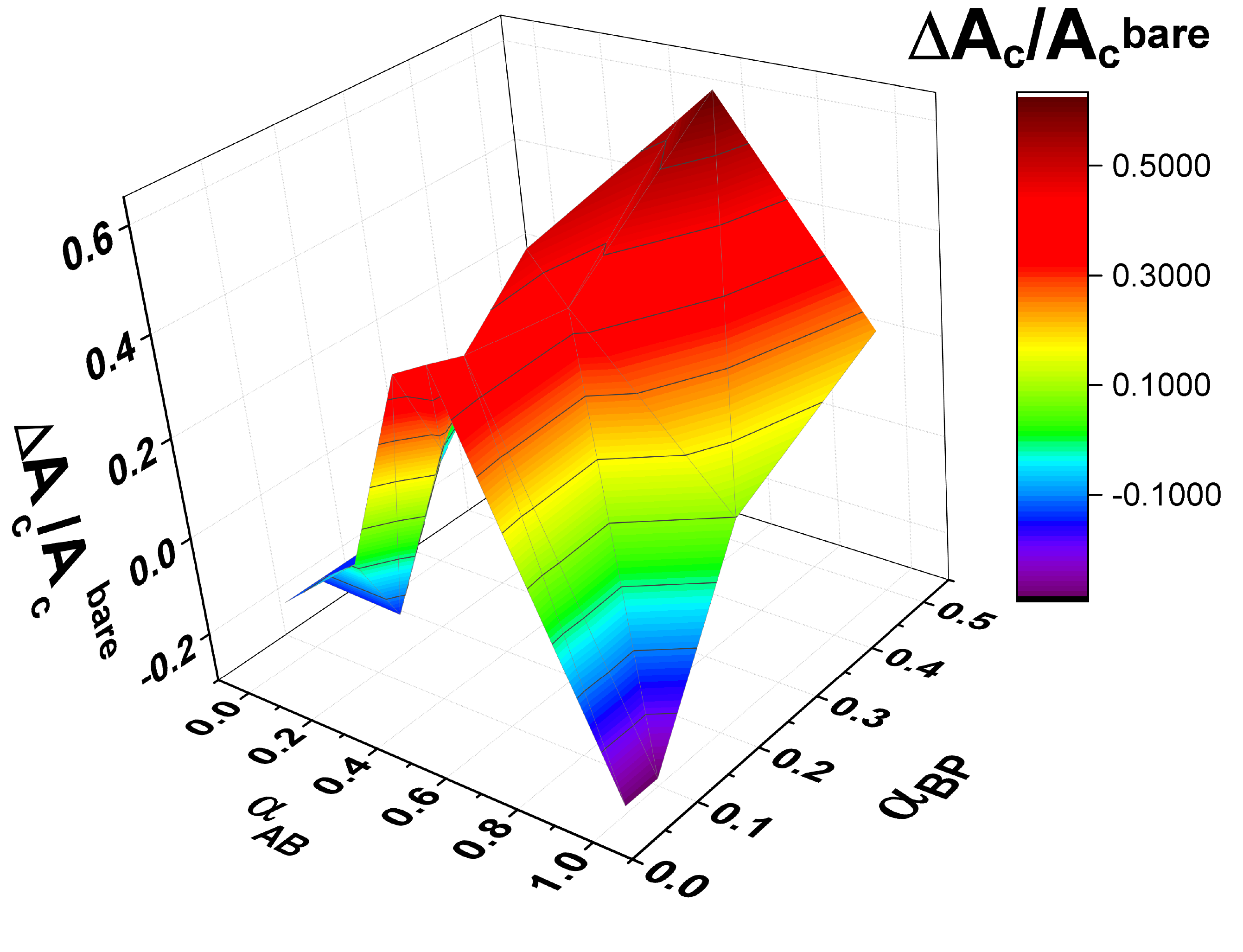}\\
\caption{Panels A,B,C,D,E: Void size distribution $P(A_h/\sAA^2)$ for varying miscibility parameter $\aAB$ and AP-preferential affinity ($0.025\leq\aAB\leq 0.5$) as indicated in the figure. Distributions obtained in absence of APs are also reported for reference. Solid lines are best fits obtained via equation \ref{Pvoid} for $\aBP=0.025$ and for bare mixtures.
Bottom panels: relative variation $\Delta c/c_{bare}$ and $\Delta A_c/A_c^{bare}$ of the exponent and the cut-off area defined in equation (\ref{Pvoid})}\label{voids-1-X}
\end{figure}
For lower $\aBP$ (Figure \ref{voids-1-X}-A,B,C,D,E) the AP addition affects much less the distributions $P(A_h/\sAA^2)$, that preserve their exponential tails. Notwithstanding this, it's worth noting that AP addition in mixtures at $\aAB=1$ and $\aAB=0.6$ give rise to contrasting effects: for the fully miscible system, an increase of the asymmetry of the AP-mixture interaction (i.e. decreasing $\aBP$) reduces the void sizes for large enough interaction asymmetry (small $\aBP$), pushing the distributions towards narrower tails; on the contrary when large irregular domains characterize the bare mixture, AP addition gives rise always to a widening of the distribution tails, reflecting the creation of larger voids. In the first case ($\aAB=1$, Figure \ref{voids-1-X}-A) the void size reduction is principally due to the fact that, by progressively breaking the interaction symmetry between particles, APs induce demixing and trigger the formation of large fluctuating domains. This suppresses, albeit weakly, large voids akin to the case of the bare mixtures at intermediate $\aAB$. In the second case ($\aAB=0.6$, Figure \ref{voids-1-X}-B), the enhanced particle adhesion due to the AP presence dominates, favouring large void formation as for the aforementioned $\aBP=1$ case. For the other cases, namely $\aAB<0.6$ (Figure \ref{voids-1-X}-C,D,E), the effect of APs on $P(A_h/\sAA^2)$, stay very weak but still detectable. The relative variation of the power-law decay exponent $\Delta c/c_{bare}$ and the cut-off area $\Delta A_c/A_c^{bare}$ are reported in figure \ref{voids-1-X} (bottom panels), where $\Delta c=(c-c_{bare})$, $\Delta A_c=(A_c-A_c^{bare})$, and $c_{bare}$ and $A_c^{bare}$ are the power-law exponent and the cut-off area extracted from the distributions in the absence of APs, respectively. We note that i) the power-law exponent varies almost to the same extent for any values of $\aAB$ by decreasing the AP-mixture interaction asymmetry (increasing $\aBP$), pointing in this case to a net reduction of the small void fraction; ii) the cut-off area variation is maximum for $\aAB=0.6$ and decreases down to negative values for $\aAB=1$ and large interaction asymmetry (low $\aBP$), confirming that the largest relative variation of the cut-off area, when APs interact preferentially with one species of particles, occurs for intermediate values of $\aAB$. This, we stress once more, is due to the fact that bare mixtures with intermediate $\aAB$ values are characterized the least number of large voids (small cut-off areas) and the addition of APs with preferential interaction with one of the two particle types (A here) gives rise to spatially localized (large) voids at the interface between the formed domains, since AP addition induces particle demixing as well as an increase of the cohesive energy in the system.  
The entire set of best-fit parameters for the data shown in figures \ref{voids-1-1} and \ref{voids-1-X} are in Table \ref{paramP} (appendix \ref{appC}).
\begin{figure}[htbp]
\includegraphics[width=8.0cm]{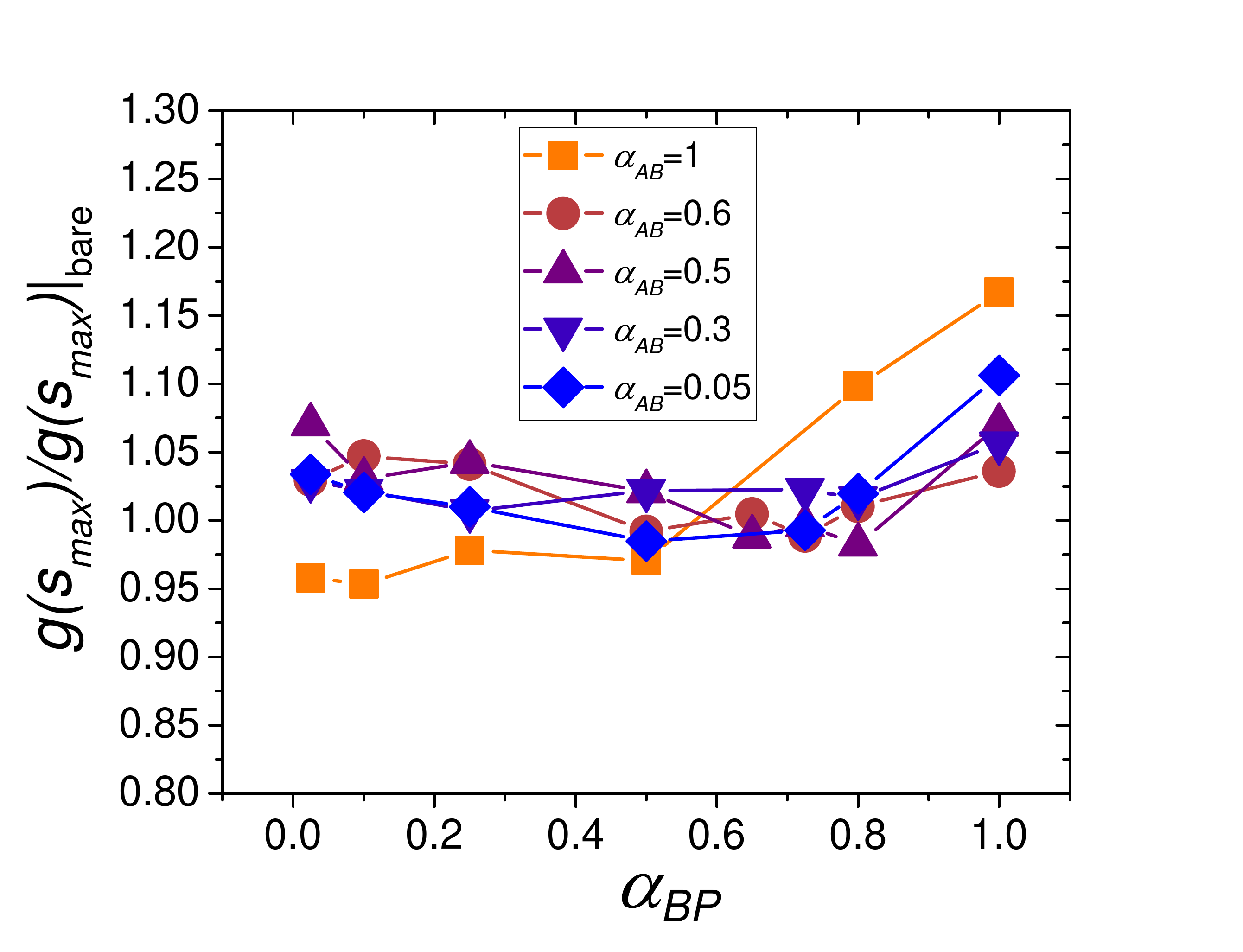}
\caption{Ratio $g(s_{max})/g(s_{max})|_{bare}$ in function of $\aBP$ for different miscibility parameters $\aAB$ as indicated in the figure. Solid lines are a guide to the eye.}\label{gmaxions}
\end{figure}
In line with the large variation of the void distribution observed for miscible systems the local structure of mixtures is most affected by the AP addition for $\aAB=1$. Figure \ref{gmaxions} shows the height of the first peak of the total pair correlation function $g(s_{max})$ normalized by its value in the absence of APs $g(s_{max})|_{bare}$. For $\aAB=1$ this quantity changes the most for varying $\aBP$ going from values larger than 1 (AP-induced condensation) for $\aBP=1,0.8$  to values lower than 1 (AP-induced fluidization) for $\aBP\leq 0.5$. In the other cases $0.05\leq\aAB\leq 0.6$ the local structure is less affected by the AP adsorption, though we may note that in this case $g(s_{max})/g(s_{max})|_{bare}$ is always larger than one for $\aBP=1$, for which the minimum value is attained at $\aAB=0.6$, where the void size distribution resulted less affected by AP addition (Figure \ref{voids-1-1}). Finally, we may note that the normalized peak height shows, albeit very weakly, a systematic non-monotonic behavior, suggesting a weak melting of the structure for intermediate values of $\aBP$. More detailed simulations would be required to reach more quantitative conclusions on this aspect.
\section{Conclusions}\label{conclusion}
In this paper we have studied the mixing-demixing transition of symmetric mixtures characterized by varying miscibility and confined to move on a sphere by using a minimal quasi-2D model. By computing the total and the partial pair correlation functions, and by employing a geodesic mixing parameter $\Xi$, we have shown that: i) demixing on 3D-spheres is a structurally smooth process, in which fully mixed and Janus-like configurations are separated by states of spatially fluctuating domains appearing for intermediate miscibility; ii) void size distributions show exponential tails preceded by a power-law decay with large void formation being hampered by large compositional fluctuations, that instead favor the appearance of small ($\sim$one-particle sized) voids. Such a scenario does not vary remarkably in the range of system sizes investigated ($R_s=11.18\sAA,7.90\sAA,5.59\sAA$). iii) Particle adsorption can give rise to both induced-mixing and induced-demixing, depending on whether mixtures are characterized by strongly fluctuating inter-domain boundaries or not. On the one hand, when coexisting domains and fluctuating interfaces characterize the fluids, uniformly adsorbed particles ($\aBP=1$) enhance miscibility, while the latter is reduced for a large preferential affinity of APs for one of the two components of the mixtures ($\aBP\ll 1$). On the other hand, for fully miscible ($\aAB=1$) or Janus-like mixtures ($\aAB\ll 1$) AP addition mainly produces particle demixing, whose extent, quantified by the normalized geodesic parameter $\Xi^*$, increases progressively with increasing miscibility, reaching its apex for $\aAB=1$ and $\aBP\ll 1$.
Void size distributions are also affected by AP addition. A transition from exponentially-tailed to fat-tailed distributions, where one single large void dominates over the others, is observed for uniform particle adsorption, namely where AP-mixture cohesion energy is maximum. The extent of such a transition also depends on the AB miscibility: fat-tailed distributions are less pronounced by large domain fluctuations occurring for intermediate $\aAB$ values, while for large enough AP-mixture interaction asymmetry, we recover exponentially-tailed distributions, with the largest tail widening and narrowing obtained respectively for an intermediate miscibility ($\aAB$=0.6) and fully miscible fluids ($\aAB$=1). Our model, albeit simple, allowed to discern under which mixing conditions a quasi 2D multicomponent fluid confined to move on a sphere is more susceptible to incur structural changes due to the interaction with an external agent. Further studies, based on the same model, will elucidate the role of mixture composition, temperature and AP concentration, \textcolor{myc}{while the effect of local curvature can be inspected by changing the confining manifold, passing from spheres to ellipsoids or tori.} We hope that our results can pave the way for more targeted experiments to investigate the role of complex molecules and particles on the mixing state of quasi-2D liquids.         

\begin{acknowledgments}
I wish to thank Dr. Simona Sennato for inspiring discussions.
We acknowledge financial support from the Agence Nationale
de la Recherche (Grant ANR-20-CE06-0030-01; THELECTRA).
\end{acknowledgments}

\section*{Data Availability Statement}
The data that support the findings of this study are all available
within the article.

\appendix
\section{Mean squared angular displacement on the sphere}\label{appA}
\begin{figure}[htbp]
\includegraphics[width=8.0cm]{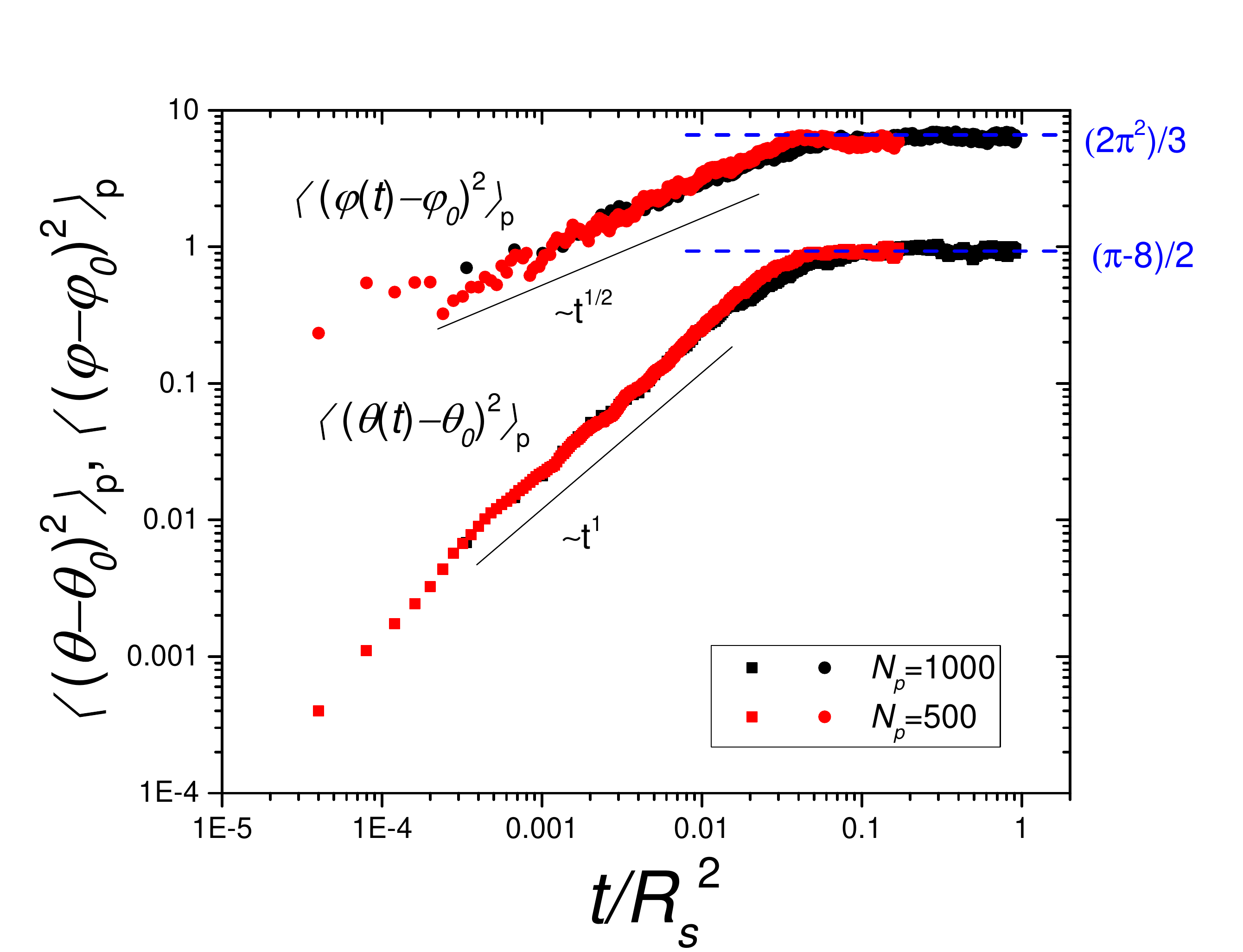}
\caption{Azimuthal and polar mean squared displacement in function of the rescaled time $t/R_s^2$ for $N_p=500$ and $N_p=1000$. Dashed horizontal lines indicate the saturation values obtained analytically (Eqs. \ref{MSDav} and \ref{MSDav2}).}\label{diffusion}
\end{figure}

We have performed long NVT simulations ($5\cdot 10^8$ and $10^9$ time steps) for the two largest system investigated ($N_p=500,1000$) and $\aAB=1$ to compute the angular mean square displacements (MSD) $\langle(\theta(t)-\theta_0)^2\rangle_p^2$ and $\langle(\varphi(t)-\varphi_0)^2\rangle_p$, and to check whether our model produces diffusive trajectories over large runs. Here $\theta(t)$ and $\varphi(t)$ represent the polar and the azimuthal angle of each particle confined on the sphere at time $t>0$, $\theta_0$ and $\varphi_0$ are their values at the initial time $t=0$, and the brackets $\langle\cdot\rangle_p$ stand for the average over all particles confined on the sphere. Figure \ref{diffusion} shows in a log-log scale both the polar and the azimuthal squared displacements, for which we observe 2 different saturation values as expected for confined dynamics. Such values can be computed analytically. The angular MSD of one Brownian particle on the sphere saturate to values that are dependent on their initial position ($\theta_0,\varphi_0$) \cite{apazaBrownianSelfdrivenParticles2017}:
\begin{align}
& \overline{(\theta(t)-\theta_0)^2}=\frac{\pi^2-4}{2}+\theta_0(\theta_0-\pi) \label{MSD}\\
& \overline{(\varphi(t)-\varphi_0)^2}=\frac{4\pi^2}{3}+\varphi_0(\varphi_0-2\pi) \label{MSD2}
\end{align}
where $\overline{(\cdot)}$ is the average over different trajectories given an initial angular position ($\theta_0,\varphi_0$). For a crowded system particles are distributed over the whole spherical surface and we must compute the saturation values of the MSD as the angular average of equations \ref{MSD} and \ref{MSD2}, i.e.:
\begin{align}
&\langle(\theta-\theta_0)^2\rangle_p=\int_0^{2\pi}d\varphi_0\int_0^{\pi}\overline{(\theta-\theta_0)^2}\sin(\theta_0)d\theta_0=\frac{\pi-8}{2} \label{MSDav}\\
&\langle(\varphi-\varphi_0)^2\rangle_p=\int_0^{\pi}\sin(\theta_0)d\theta_0\int_0^{2\pi}\overline{(\varphi-\varphi_0)^2}d\varphi_0=\frac{2\pi^2}{3} \label{MSDav2}
\end{align}
The values obtained are in very good agreement with our simulations. We also note that the polar MSD is a linear function of $t$ for small lag times in accordance with recent theoretical works \cite{apazaBrownianSelfdrivenParticles2017,castaneda-priegoBrownianMotionFree2013}, while the azimuthal MSD shows a subdiffusive behavior $\langle(\varphi-\varphi_0)^2\rangle_p\sim t^{0.47\pm0.04}$, that, up to our knoweldge, has not been observed so far. $\varphi(t)$ in a spherical set of coordinates isn't in fact a variable satisfying the Langevin equation and we attribute this feature to the space-dependent azimuthal diffusivity \cite{apazaBrownianSelfdrivenParticles2017}: particles close to the pole $\theta=0$ move faster in $\varphi$ with respect to particles moving around the equator $\theta=\pi/2$ for a fixed geodesic displacement. Sub-diffusive motions are often observed in experiments and simulations in systems characterized by space-dependent diffusion constants \cite{cherstvyParticleInvasionSurvival2014}. This aspect, interesting \emph{per se}, goes beyond the scope of this paper and surely deserves a deeper investigation.
Finally the displacements obtained for the two systems confined onto spheres of different radius $R_s$ superimpose when a rescaled time $t/R^2$ is considered, in agreement with the general solution of the diffusion equation for spherical Brownian motion. \cite{yosidaBrownianMotionSurface1949,castaneda-priegoBrownianMotionFree2013}.     
\section{Fit of the geodesic mixing parameter $\Xi(\aAB)$}\label{appB}
 We extracted the critical mixing parameter $\aAB^c$ from fits
 of the geodesic mixing parameter $\Xi(\aAB)$ through the phenomenological function
 \begin{multline}\label{fitxi}
  \Xi(\aAB)=\aAB^0+\Delta\alpha_{\scriptscriptstyle AB}\tanh\left[(\psi(\aAB-\aAB^c)\right]\\+A(\aAB-\aAB^c)+B(\aAB-\aAB^c)^2  
 \end{multline}\\
We obtain the critical miscibility parameters $\aAB^c$ reported in Table \ref{criticalalpha}.
\begin{table}[h!]
    \begin{tabular}{|c||c|}
        \hline
        $N_p$                           & $\aAB^c$ \\ \hline
        250                             & 0.40 $\pm$ 0.02 \\
        500                             & 0.43 $\pm$ 0.04 \\
        1000                            & 0.49 $\pm$ 0.01 \\ 
        \hline
    \end{tabular}
    \caption{Critical miscibility parameter $\aAB^c$ obtained by fitting $\Xi(\aAB)$ via equation \ref{fitxi}.}\label{criticalalpha}
\end{table}
\section{Best-fit parameters for $P(A_h/\sAA^2)$}\label{appC} 
We report below (Table \ref{paramP}) the best-fit parameters obtained by fitting the void size distributions shown in figure \ref{voids-1-1} and \ref{voids-1-X} via equation (\ref{Pvoid}).
\begin{table}[h!]\label{paramP}
    \begin{tabular}{|c|c||c|c|c|c|c|}
        \hline
        $\aAB$ & $\aBP$ & $P_0$ & $a$ & $b$ & $c$ & $A_c/\sAA^2$\\\hline
          1    &    1   &  $2.95\cdot 10^{-5}$ & 1.95 & 4.68 &  2.24  & $\infty$\\\hline
          0.6    &   1   &  0    &  1.25  &  3.28  &  2.33  & $\infty$ \\\hline     
          0.5    &   1   &  0    &  0.28  &  6.04 &  2.48  &      $\infty$ \\\hline
          0.3    &   1   &  0    &  1.53  &  4.08  &  2.12  &      $\infty$ \\\hline
          0.05    &   1   &  $4.5\cdot 10^{-5}$ & 1.40 & 3.64 & 2.32 & $\infty$\\\hline
          1    &    0.5   &  0    &  1.35  &  3.52  &  2.10  &      9.17 \\\hline
          0.6    &   0.5   &  0    &  1.40  &  3.56  &  2.07  &     9.24  \\\hline
          0.5    &   0.5   &  0    &  1.34  &  3.72  &  2.01  &      8.68 \\\hline
          0.3    &   0.5   &  0    &  1.35  &  3.56  &  1.97  &      6.32 \\\hline
          0.05    &   0.5   &  0    &  1.44  &  3.88  &  1.95  &      8.92 \\\hline
         1    &    0.25   &  0    &  1.38  &  3.6  &  2.06  &      7.66 \\\hline
          0.6    &   0.25   &  0    &  1.33  &  3.48  &  1.99  &      5.73 \\\hline
          0.5    &   0.25   &  0    &  1.57  &  4.24  &  1.88  &      7.45 \\\hline
          0.3    &   0.25   &  0    &  1.52  &  4.13  &  1.90  &      7.75 \\\hline
          0.05    &   0.25   &  0    &  1.65  &  4.52  &  1.79  &      6.00 \\\hline
          1    &    0.1  &  0    &  1.41  &   3.65 & 2.02  &      5.50 \\\hline
          0.6    &   0.1   &  0    &  1.76  &  4.56  &  1.89  &      6.00 \\\hline
          0.5    &   0.1   &  0    &  1.70  &  4.56  &  1.78  &      5.00 \\\hline
          0.3    &   0.1   &  0    &  1.74  &  4.72  &  1.74  &      5.00 \\\hline
          0.05    &   0.1   &  0    &  1.84  &  5.08  &  1.74  &      6.00 \\\hline
          1    &    0.025   &  0    &  1.62  &  4.22  &  1.91  &      5.50 \\\hline
          0.6    &   0.025   &  0    &  1.55  &  4.14  &  1.90  &      6.25 \\\hline
          0.5    &   0.025   &  0    &  1.58  &  4.24  &  1.88  &      6.50 \\\hline
          0.3    &   0.025   &  0    &  1.75  &  4.80  &  1.76  &      5.75 \\\hline
          0.05    &   0.025   &  0    &  1.49  &  4.16  &  1.80  &      6.00 \\
        \hline
    \end{tabular}\caption{Best fit parameters obtained via equation \ref{Pvoid} for the void size distributions shown in figures \ref{voids-1-1} and \ref{voids-1-X}.}\label{paramP}
\end{table}
\textcolor{myc}{\section{Sampling time}\label{appD} 
The sampling time ($\Delta t_s=5\cdot10^{5}$ time steps) has been chosen by computing the (spherical) self-intermediate scattering function for the bare mixture with the lowest energy ($N_p=1000$ and $\aAB=1$) as defined in \cite{singhCooperativelyRearrangingRegions2020}}:

\begin{equation}\label{fcorr}
\textcolor{myc}{F_s(k,t)=\frac{1}{N_p}\sum\limits_{j=1}^{N_p}\left\langle P_{kR}\left[\cos\left(\frac{\Delta s_j(0,t)}{R}\right)\right]\right\rangle}
\end{equation}
 
\textcolor{myc}{Where, $P_{kR}$ is the Legendre polynomial with {kR} being rounded-off to the nearest integer, 
$k^{-1}$ is the wavevector corresponding to the geodesic distance equal to one particle diameter and $\Delta s_j(0, t)$ is the geodesic displacement of particle $j$ over time $t$. We obtain the scattering function shown in figure \ref{fcorr} that is well fitted by a single exponential decay with a relaxation time of $8.29 \cdot 10^4$  time steps that is about 6 time less than the fixed $\Delta t_s$.}
\begin{figure}[htbp]
\includegraphics[width=8cm]{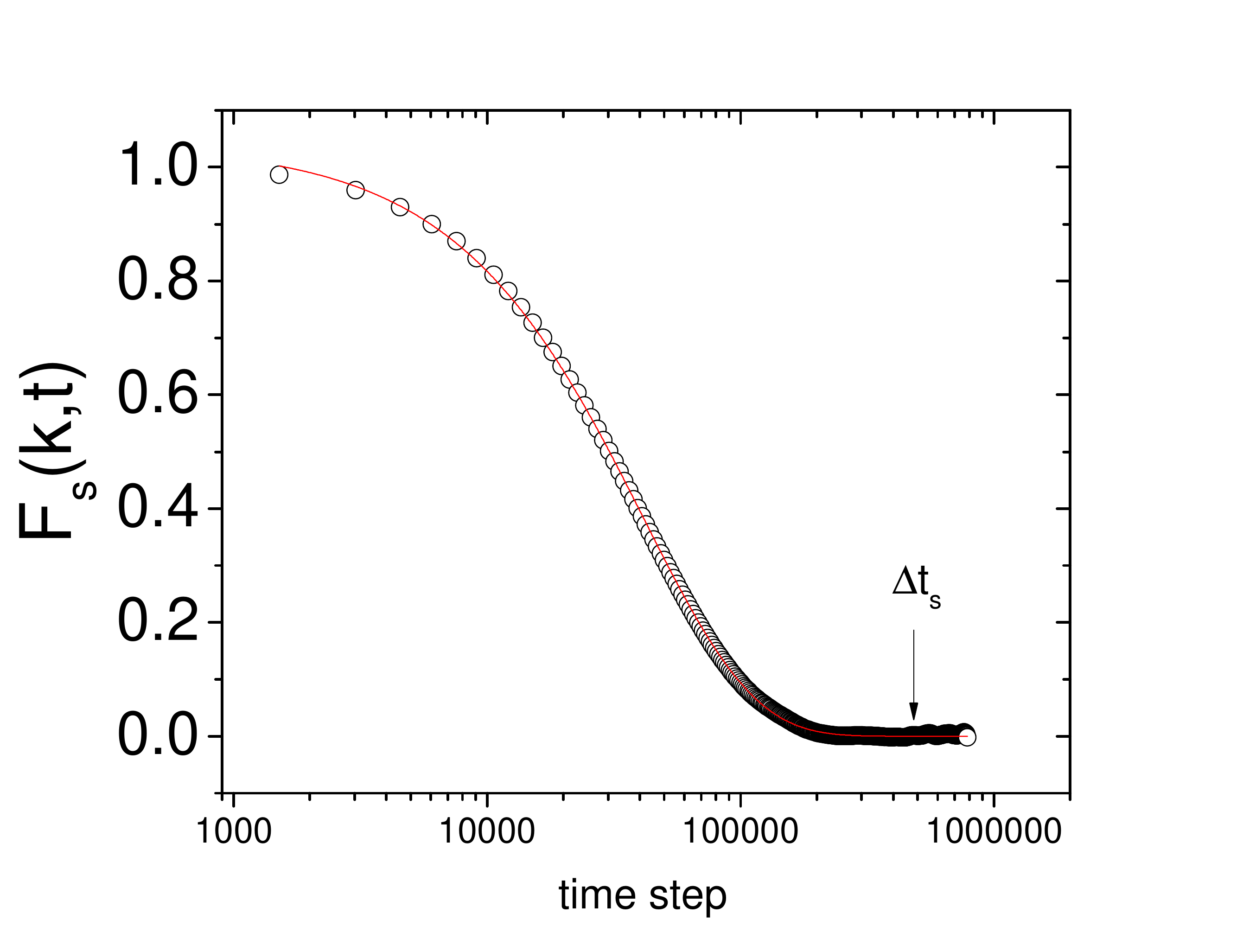}
\caption{\textcolor{myc}{Self-intermediate scattering function at a wavevector corresponding to a geodesic distance equal to one particle diameter for $N_p=1000$ and $\aAB=1$.}}\label{fcorr}
\end{figure}


\bibliography{Paper-LJsphere}

\end{document}